\documentclass[hidelinks, conference]{IEEEtran}
\IEEEoverridecommandlockouts
\PassOptionsToPackage{bookmarks={false}}{hyperref}

\usepackage[a-1b]{pdfx}
\usepackage{color}
\usepackage{url}
\usepackage[utf8]{inputenc}
\usepackage[T1]{fontenc}
\usepackage{amsthm}
\usepackage{amsmath}
\usepackage{amsthm}
\usepackage{amssymb}
\usepackage{graphicx}
\usepackage{epstopdf}
\usepackage{wrapfig}
\usepackage{algorithmic}
\usepackage{dirtytalk}
\usepackage{algorithm}
\usepackage{mathrsfs}
\usepackage{float}
\usepackage{mathtools}
\usepackage{tabulary}
\usepackage{booktabs}
\usepackage{caption}
\usepackage{subfig}
\usepackage{xprintlen}
\usepackage{multirow}

\graphicspath{{./}{./files/figures_eps/}}

\PassOptionsToPackage{bookmarks={false}}{hyperref}

\def\BibTeX{{\rm B\kern-.05em{\sc i\kern-.025em b}\kern-.08em
    T\kern-.1667em\lower.7ex\hbox{E}\kern-.125emX}}

\newcommand{\comment}[1]{ }

\newcommand\subparagraph{%
  \@startsection{subparagraph}{0}
  {\parindent}
  {0ex \@plus 0ex \@minus 0ex}
  {-1em}
  {\normalfont\normalsize\bfseries}}
\makeatother

\usepackage{titlesec}
\titlespacing*{\section}
{0pt}{1.2ex}{0ex} 
\titlespacing*{\subsection}
{0pt}{2.2ex}{0ex}  
\titlespacing*{\subsubsection}
{0pt}{0ex}{0ex} 

\begin{document}

\title{On the Impact of the Hardware Warm-Up Time on Deep Learning-Based RF Fingerprinting}

\author{Abdurrahman Elmaghbub, Vincent Immler and Bechir Hamdaoui~\\
 Oregon State University, Corvallis, OR, USA ~\\ 
 \{elmaghba,vincent.immler,hamdaoui\}@oregonstate.edu
 \thanks{This work is supported in part by Intel/NSF Award No. 2003273.}
}

\maketitle
\thispagestyle{plain}
\pagestyle{plain}

\begin{abstract}
Deep learning-based RF fingerprinting offers great potential for improving the security robustness of various emerging wireless networks. Although much progress has been done in enhancing fingerprinting methods, the impact of device hardware stabilization and warm-up time on the achievable fingerprinting performances has not received adequate attention. As such, this paper focuses on addressing this gap by investigating and shedding light on what could go wrong if the hardware stabilization aspects are overlooked. Specifically, our experimental results show that when the deep learning models are trained with data samples captured after the hardware stabilizes but tested with data captured right after powering on the devices,
the device classification accuracy drops below 37\%.
However, when both the training and testing data are captured after the stabilization period, the achievable average accuracy exceeds 99\%, when the model is trained and tested on the same day, and achieves 88\% and 96\% when the model is trained on one day but tested on another day, for the wireless and wired scenarios, respectively.
Additionally, in this work, we leverage simulation and testbed experimentation to explain the cause behind the I/Q signal behavior observed during the device hardware warm-up time that led to the RF fingerprinting performance degradation. Furthermore, we release a large WiFi dataset, containing both unstable (collected during the warm-up period) and stable (collected after the warm-up period) captures across multiple days. Our work contributes datasets, explanations, and guidelines to enhance the robustness of RF fingerprinting in securing emerging wireless networks.
\end{abstract}

\begin{IEEEkeywords}
Device fingerprinting, hardware warm-up 
and stabilization, carrier frequency offset, oscillator impairments.
\end{IEEEkeywords}

\section{Introduction}
\label{sec:into}
Deep learning (DL)-based RF fingerprinting emerges as a compelling physical-layer security solution \cite{hamdaoui2020deep, sankhe2019oracle, jian2020deep}, enabling automated device identification and authentication through the extraction of device-specific fingerprints from I/Q samples collected from the received RF signals.
These fingerprints exist due to inherent device hardware imperfections introduced during the manufacturing of circuitry components in the RF signal path~\cite{elmaghbub2020widescan, elmaghbub2020leveraging, rajendran2022rf, smaini2012rf}. Because such hardware imperfections are random and differ from device to device, they create per-device-unique signatures that can be leveraged to identify and distinguish devices from one another. With the power of deep learning, I/Q data-driven fingerprinting technology brings then great potential for enabling a wide range of network security services, including safeguarding against unauthorized access to critical data, uncovering the identity of malicious activities in the network, and fortifying the overall security of interconnected cyber-physical infrastructures \cite{jagannath2022comprehensive}.

Despite the rich amount of work that has been done on improving the robustness and performance of DL-based RF fingerprinting methods \cite{jagannath2022comprehensive,basha2023channel}, it is our understanding that the effect of hardware stabilization and warm-up time (i.e., the time needed for the circuit components to reach their stable conditions after being powered on) has not received adequate attention. Our objective in this work is then to address this gap and shed some light on what could go wrong if the hardware stabilization aspects are overlooked when developing and evaluating RF data-driven device fingerprinting methods. 

Our recent studies \cite{elmaghbub2021,adl} have indicated a significant degradation in the RF fingerprinting performance when the training data and testing data are collected even just a few minutes apart from one another if measurements are taken before the hardware is stabilized. These studies have prompted us to investigate the link between the observed drop in performance and the unstable behavior of RF hardware components (e.g. local oscillators) during the warm-up period. For this, we collected WiFi data, using a testbed of 15 Pycom (IoT) devices, at various intervals within the initial 20 minutes after powering on the devices to closely monitor the behavior of the I/Q signal during the warm-up period and thoroughly assess its impact on the performance of I/Q data-driven fingerprinting methods. Remarkably, our findings show that when training the fingerprinting learning model with data captured after the hardware is stabilized (i.e., after about 12 minutes from device activation) and testing it with data captured right after activating the devices, the device classification accuracy drops significantly, to about an average of 30\% and 20\% for the wireless and wired setups, respectively.
This dramatic drop when having a 12-minute gap between the training and testing data collection times indicates substantial signal variations occurring during the warm-up period, rendering the classifier unable to accurately recognize the distinct characteristics of the signals.
On the other hand, our findings also show that when postponing the acquisition of the training and testing data until after the hardware is stabilized, the classification accuracy for the wireless and wired setups reaches respectively an average of 99.3\% and 99.7\%, when the model is trained and tested on the same day, and 88\% and 96\%, when the model is trained on one day but tested on another day. 
This significant improvement in the cross-day accuracy when considering ``stable captures" suggests that the fingerprints remain stable over time once the warm-up period is over. 

Our research findings provide further understanding and explanation of the time-sensitivity of deep learning-based RF fingerprinting methods that recent studies have reported~\cite{elmaghbub2021, adl, hamdaoui2022deep,al2020exposing,gaskin2022tweak,hanna_wisig_2022,hamdaoui2023deep}.
For instance, when the fingerprinting learning model is trained on data captured on one day but tested on data captured on a different day, a notable decline of more than 60\% in performance was observed in~\cite{elmaghbub2021,al2020exposing}. While it is widely believed that the wireless channel is the main contributor to such a performance drop~\cite{al2020exposing}, the remarkable instability of the fingerprinting accuracy during hardware warm-up that our work reveals made us speculate and question such a belief. In contrast, our findings suggest that the real cause of the fingerprinting accuracy degradation across time may not be the wireless channel, but rather the instability of the hardware during warm-up time. The similar performance drop observed in wired scenarios \cite{elmaghbub2021, al2020exposing} where the influence of the wireless channel is minimal or when the gap between the training and testing data collection times is set to only 5 minutes \cite{adl} are both in effect a confirmation that the significant performance drop reported in various cross-day studies~\cite{elmaghbub2021,hanna_wisig_2022,hamdaoui2023deep} could be due to the lack of hardware stabilization during the warm-up phase.

The contributions of our work are summarized as follows: 
\begin{itemize}
    \item We study and explain, through thorough testbed experimentation, the impact of the hardware warm-up time on the accuracy performance of deep learning-based RF fingerprinting when used for device classification.
    
    \item We show that deep learning models that are trained with data captured after the stabilization period (approximately 12 minutes from the activation of the tested devices) but tested with data captured at the beginning of the warm-up period of the same power cycle fail to maintain a satisfactory performance as their average device classification accuracies drop as low as 30\% and 20\% for the wireless and wired setups, respectively. 
    
    \item We demonstrate that when the training and testing data have been captured after hardware stabilization, same-day classification accuracy exceeds 99\% for both the wireless and wired setups. Furthermore, even when the learning model is trained on data collected on one day and tested on data collected during another day, the classification accuracy remains high at 88\% and 96\% for the wireless and the wired scenarios, respectively.  
    
    \item We demonstrate and explain the root cause behind the time-domain I/Q signal behavior that was observed during the device hardware warm-up and stabilization time and that led to a dramatic degradation in the device classification accuracy of the RF fingerprinting.
    
    
    \item We release massive WiFi datasets of ``stable" captures (collected after the warm-up period) from 15 Pycom devices on multi-day wireless and wired scenarios. These can be downloaded from NetSTAR Lab at 
\href{https://research.engr.oregonstate.edu/hamdaoui/datasets/}{\color{blue}{http://research.engr.oregonstate.edu/hamdaoui/datasets}}.

\end{itemize}

The rest of the paper is organized as follows. We begin by studying the impact of hardware warm-up on I/Q signal behavior in Section~\ref{sec:stability} and the RF fingerprinting accuracy thereof in Section~\ref{sec:effect}. We then present the testbed and experimental setup in Subsection~\ref{sec:setup} and the performance result evaluation and analysis in Subsection~\ref{sec:results}. 
Finally, we explain through simulations and experimentation the cause behind the observed time-domain I/Q signal behavior in Section~\ref{sec:inaccuracy} and conclude the paper in Section~\ref{sec:conc}.

\section{The Behavior of I/Q Signals During Hardware Warm-Up and Stabilization Period}
\label{sec:stability}

In order to investigate the impact of the device hardware warm-up and stabilization period on I/Q data-driven deep learning-based RF fingerprinting, it is important to first understand how the time-domain I/Q signal data behaves during the hardware warm-up period. For this, we randomly selected four Pycom devices (hereafter labeled Device A, Device B, Device C, and Device D) from the 15 devices used in our testbed (more details on this testbed are provided later) and closely monitored the behavior of their received I/Q signals during the initial 20 minutes following device activation. This involved capturing 802.11b WiFi packets transmitted by the devices using a USRP B210 receiver at 45MSps sampling rate. 

\subsection{I/Q Signal Behavior Observed Over Time}
\begin{figure}
\centering
    \begin{minipage}{\linewidth} 
        \centering 
        \subfloat[The I component of the WiFi packet]{%
        \includegraphics[width=\linewidth,height=.3\linewidth]{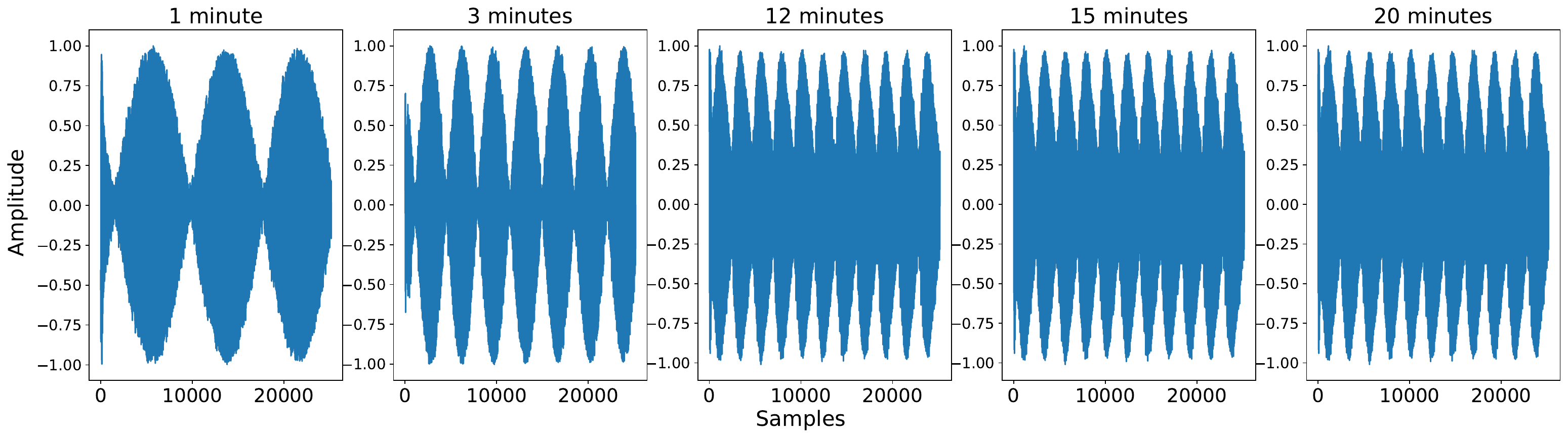}
        \vspace{-0.4in}
        \label{dev6_I}}\\ 
        \subfloat[The Q component of the WiFi packet]{%
        \includegraphics[width=\linewidth,height=.3\linewidth]{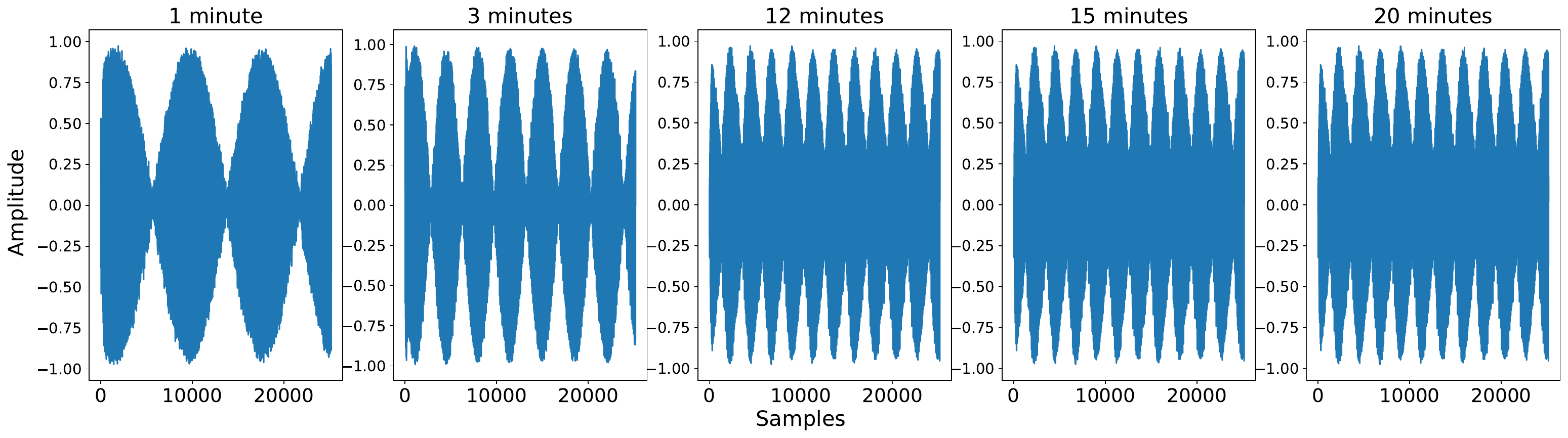}
        \label{dev6_Q}}
        \caption{I/Q signal behavior of Device A observed at different times during the device warm-up period.} 
        \label{dev11} 
    \end{minipage}
    \vfill \vspace{0.2in}
    \begin{minipage}[t]{\linewidth} 
        \centering 
         \subfloat[The I component of the WiFi packet]{%
        \includegraphics[width=\linewidth,height=.3\linewidth]{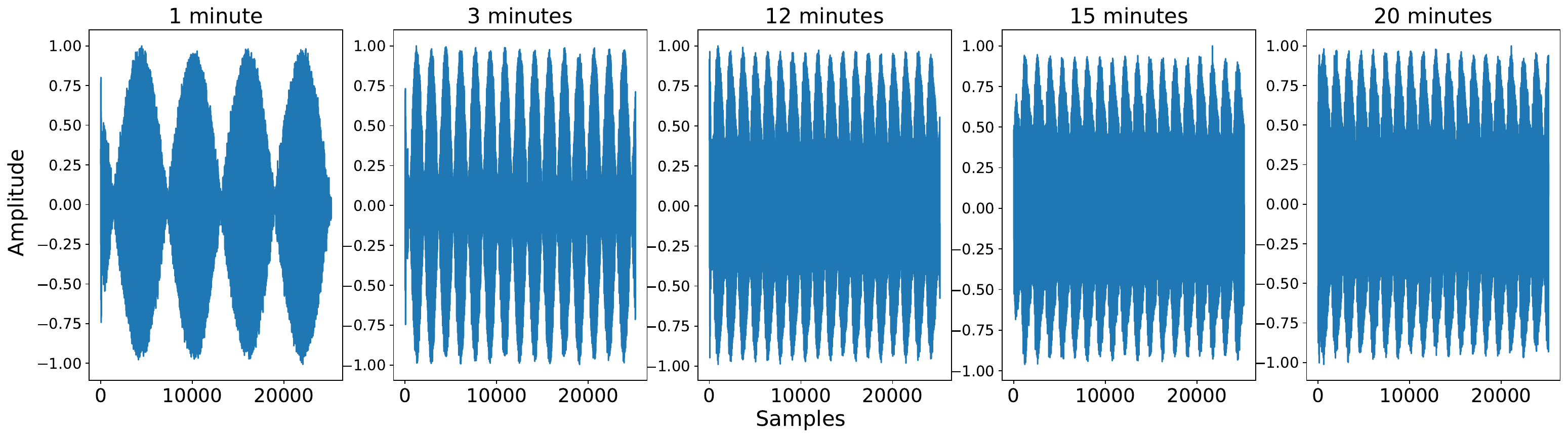} 
        \label{dev10_I}}\\ 
        \subfloat[The Q component of the WiFi packet]{%
        \includegraphics[width=\linewidth,height=.3\linewidth]{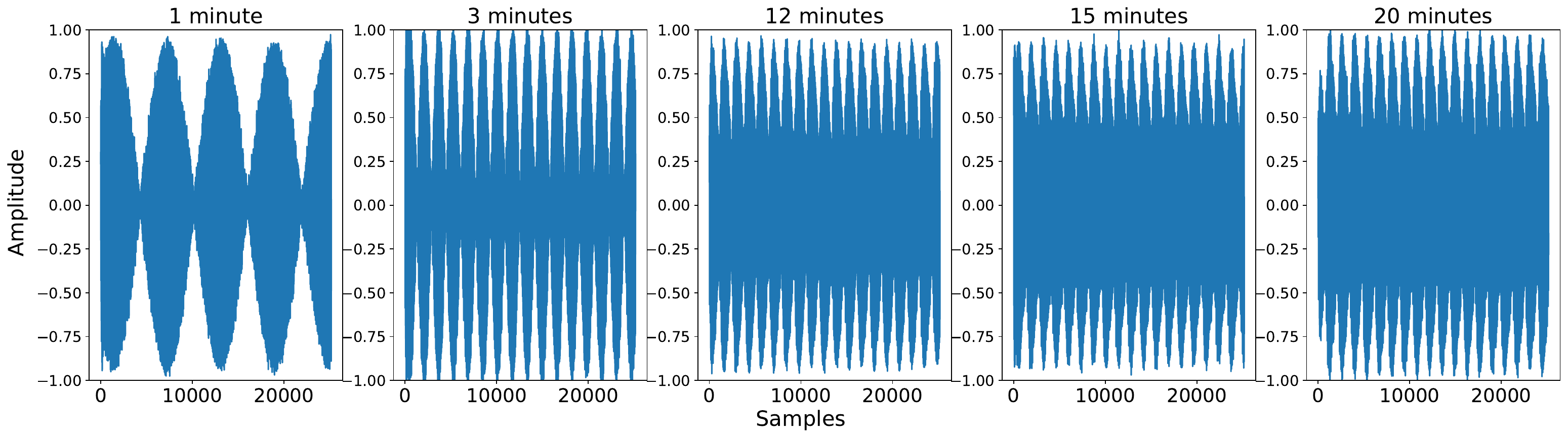}
        \label{dev10_Q}}
        \caption{I/Q signal behavior of Device B observed at different times during the device warm-up period.} 
        \label{dev10} 
    \end{minipage}
\end{figure}

We show in Fig.~\ref{dev11} the time-domain I/Q signal captured on Device A at different times during the device warm-up period; i.e., the two figures on the far-left correspond to the signal captured one minute from when the device was powered on, the two figures on the far-right correspond to the signal captured 20 minutes from when the device was powered on, and so on.
A few important observations can be drawn from this figure. 
First, observe that the shapes of the I and Q signals do change over time as the device hardware warms up, 
with the frequency of the Envelopes\footnote{The Envelope of an oscillating signal is the smooth boundary function that outlines the extremes of the signal (\cite{johnson2011software}, Appendix C).} of the signals increasing over time; i.e., the number of 'humps' in the signal's Envelope increases as the device warms up.
Second, observe that both the I and Q envelopes converge after some time. In our experiments, they converged around 12 minutes---note that even though we monitored these shapes at each minute from activation to detect when they start to converge, we only show them here after they converged; i.e., at minutes 12, 15 and 20. Our experiments show then that the hardware of these Pycom devices stabilizes at about 12 minutes from device activation.
A third observation we make here is that the envelopes of the I and Q signal components vary in the opposite direction of one another; i.e., shifted by 180 degrees.

\begin{figure}
\centering
    \begin{minipage}{\linewidth}
         \centering 
         \subfloat[The I component of the WiFi packet]{%
        \includegraphics[width=\linewidth,height=.3\linewidth]{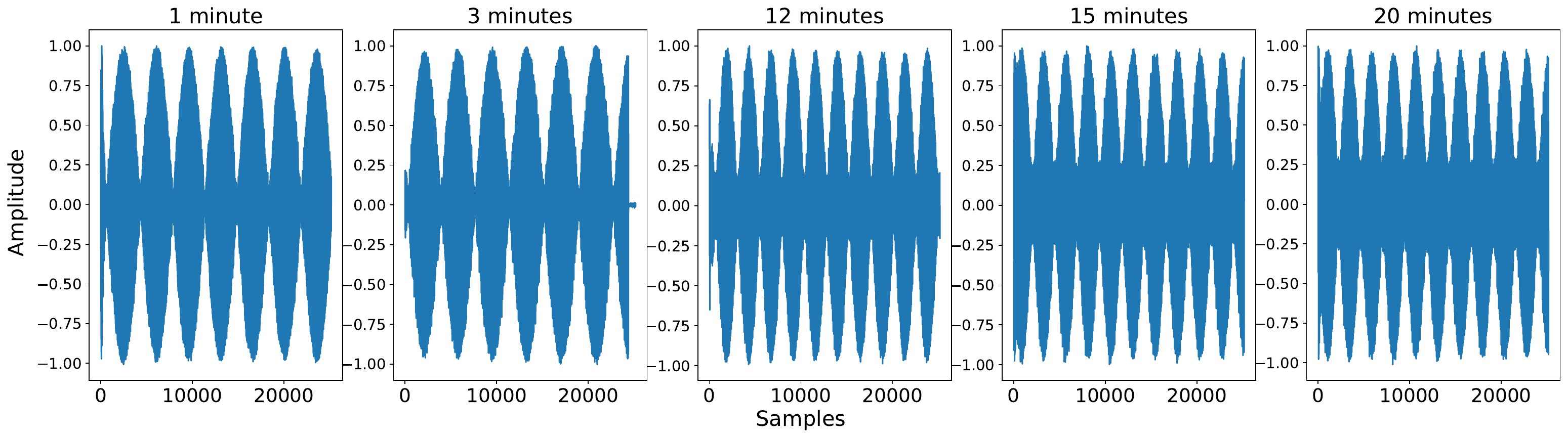}
        \label{dev11_I}}\\ 
        \subfloat[The Q component of the WiFi packet]{%
        \includegraphics[width=\linewidth,height=.3\linewidth]{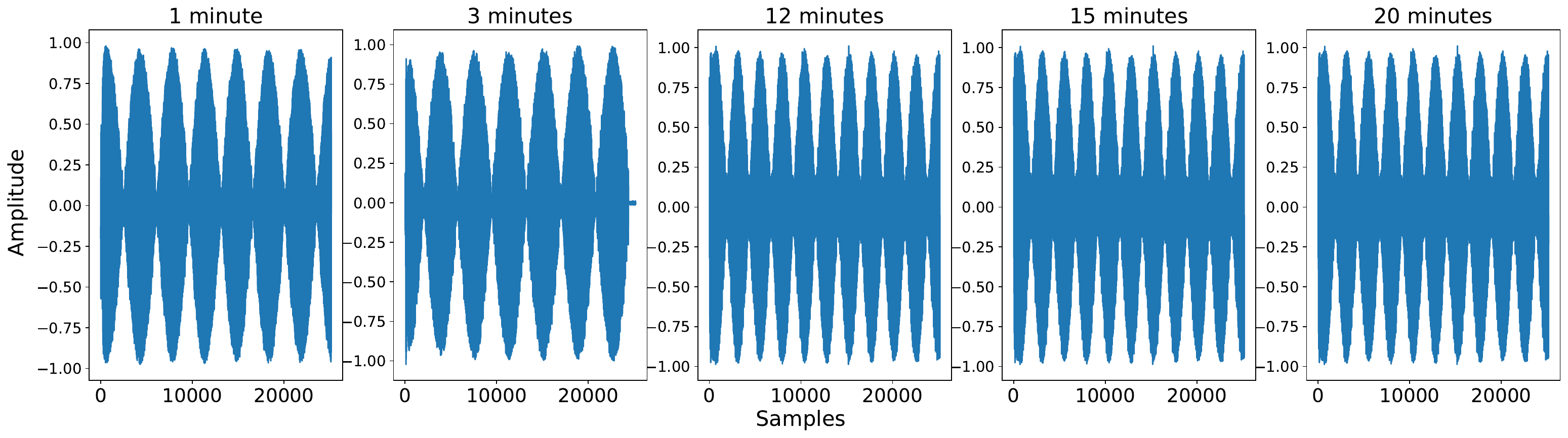}
        \label{dev11_Q}}
        \caption{I/Q signal behavior of Device C observed at different times during the device warm-up period.} 
        \label{dev6} 
    \end{minipage}
    \vfill \vspace{0.2in}
    \begin{minipage}[t]{\linewidth}
       \centering 
         \subfloat[The I component of the WiFi packet]{%
        \includegraphics[width=\linewidth,height=.3\linewidth]{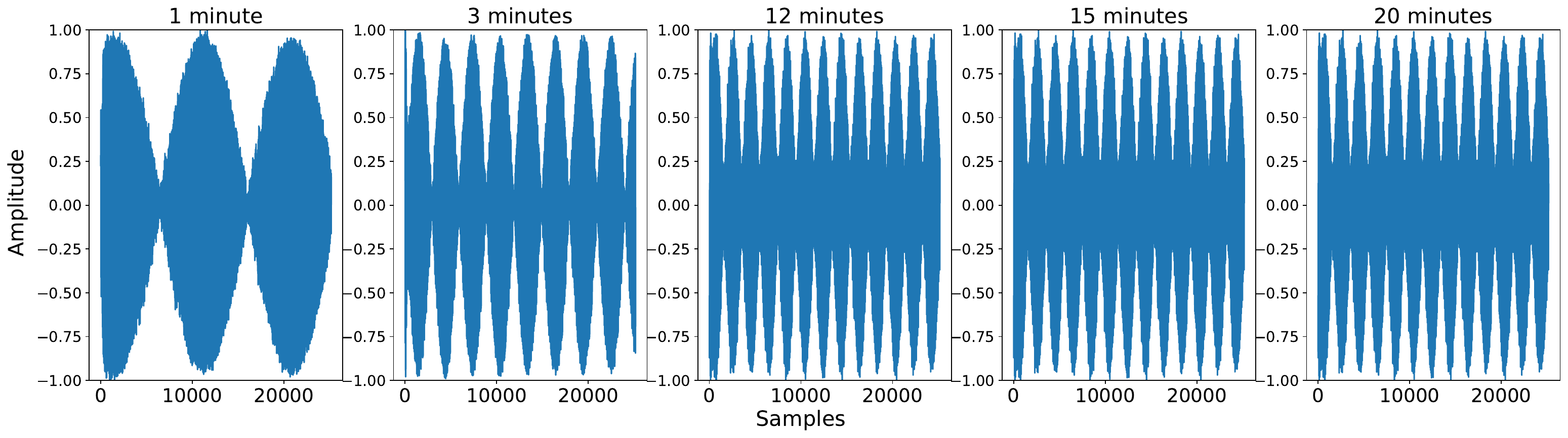} 
      \label{dev7_I}}\\ 
       \subfloat[The Q component of the WiFi packet]{%
        \includegraphics[width=\linewidth,height=.3\linewidth]{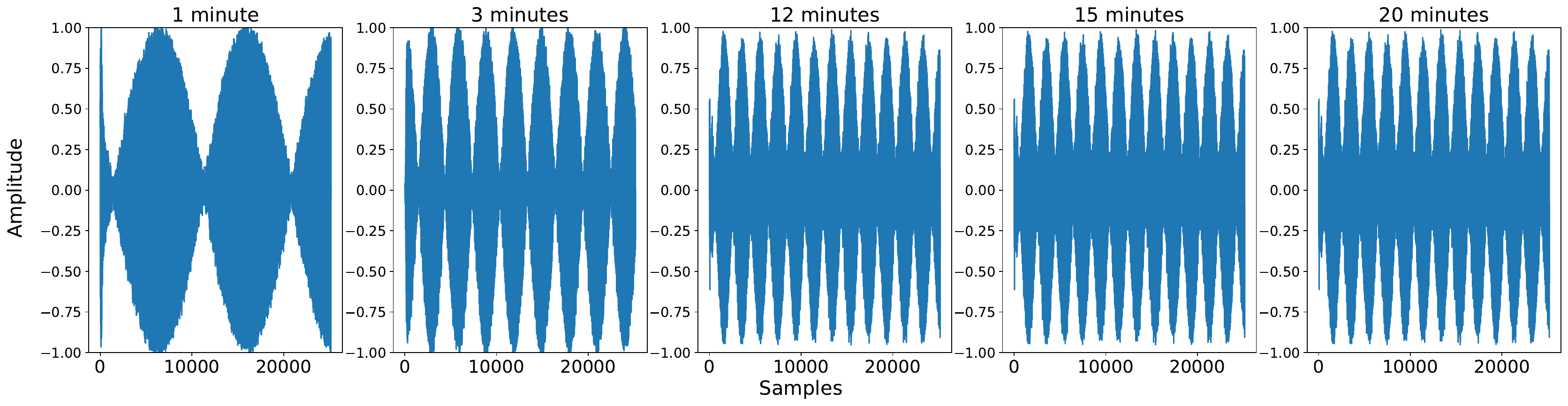}
       \label{dev7_Q}}
        \caption{I/Q signal behavior of Device D observed at different times during the device warm-up period.} 
       \label{dev7} 
  \end{minipage}
\end{figure}

To see whether these observed trends are consistent across different devices, we collected and monitored the behavior of the I/Q signals received from all other 15 testbed devices during the hardware warm-up period. Although shown for four devices only in the paper, Device A (Fig.~\ref{dev11}), Device B (Fig.~\ref{dev10}), Device C (Fig.~\ref{dev6}), and Device D (Fig.~\ref{dev7}), our experimental results confirm that the reported trends are observed across all devices, although each device exhibits slightly different (the initial as well as the more stabilized) envelopes, as can be observed in Figs.~\ref{dev11},~\ref{dev10},~\ref{dev6}, and~\ref{dev7} for Devices A, B, C and D.

\begin{figure}
\centering
    \begin{minipage}[t]{\linewidth} 
    \centering 
    \subfloat[After 1 minute from device activation]{%
    \includegraphics[width=\linewidth,height=.3\linewidth]{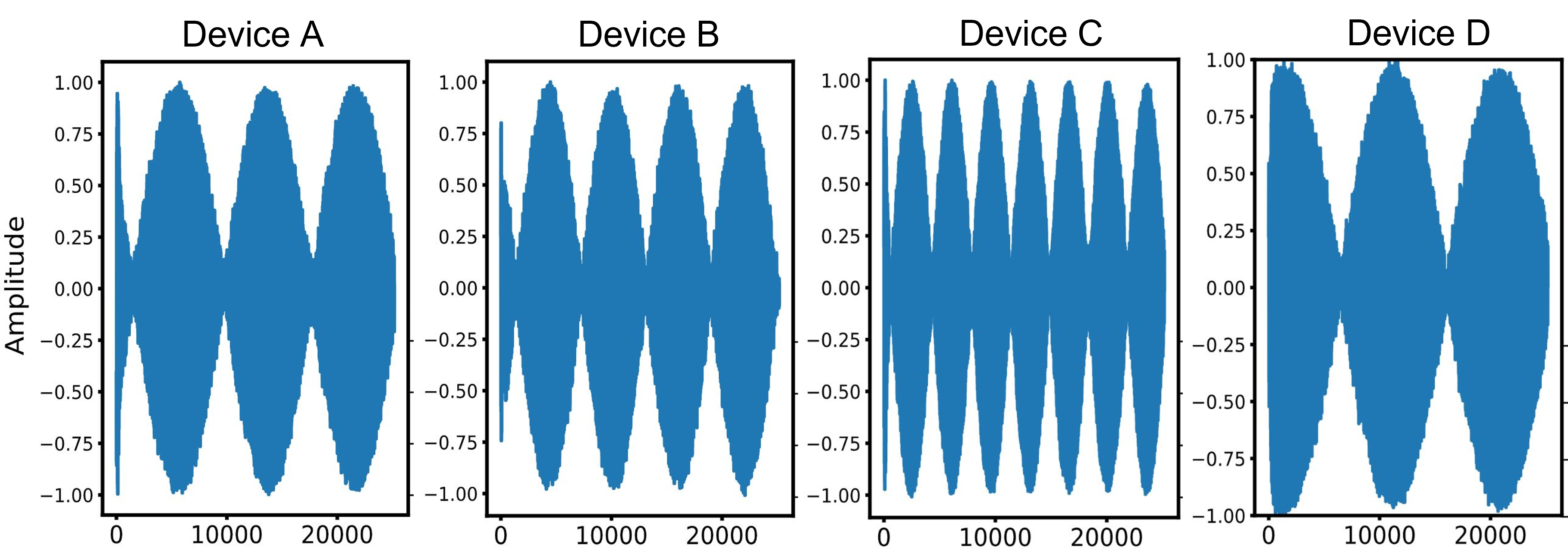} 
     \label{all-devices-1min}}\\ 
    \subfloat[After 12 minutes from device activation]{%
    \includegraphics[width=\linewidth,height=.3\linewidth]{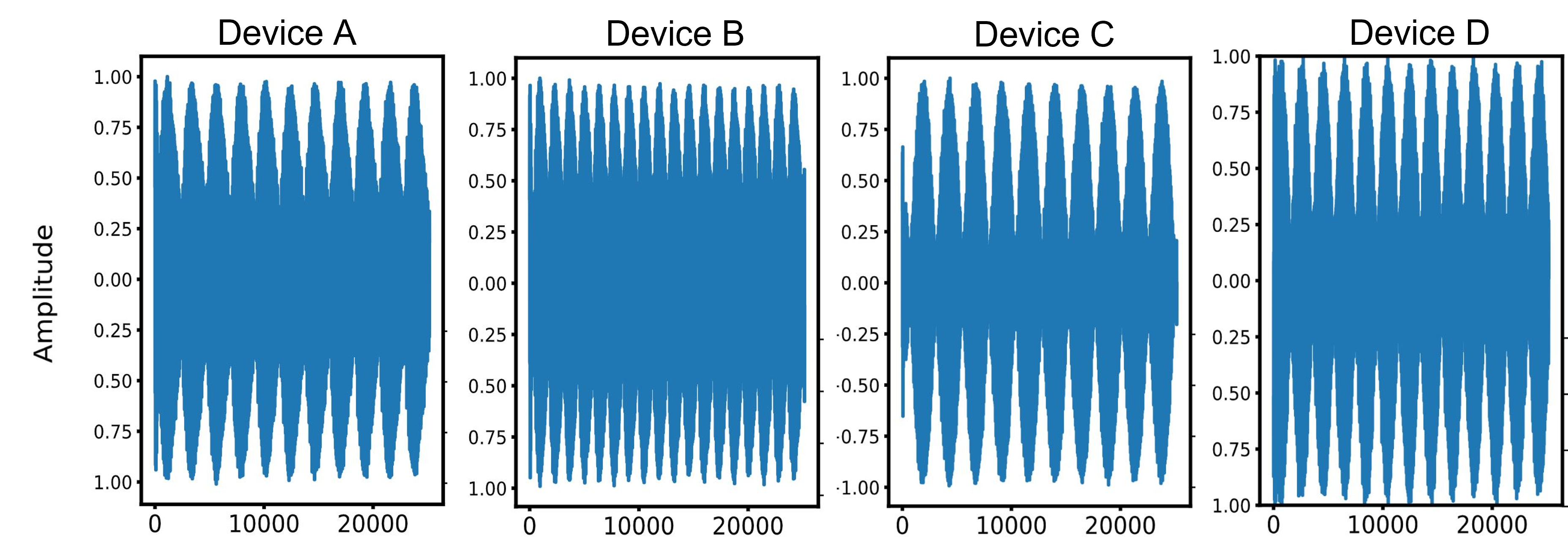}
    \label{all-devices-12min}}
    \caption{The I component of the I/Q signal.} 
    \label{all-devices-powerup} 
    \end{minipage}
\end{figure}

\subsection{I/Q Signal Behavior Observed Across Devices}
Now that we've monitored and analyzed how the I/Q signal behaves over time during the hardware stabilization period, in this section, we turn our attention to show how the signal behaves across different devices when captured at the same time from when the device is activated.
To illustrate this, we again present in Figs.~\ref{all-devices-1min} and~\ref{all-devices-12min} the I component of the I/Q signal captured from the 4 randomly selected devices respectively at 1 minute and 12 minutes from when the devices were activated. 
%
The results clearly demonstrate that right after activation (i.e., after 1 minute from power on), the number of 'humps' in the signals' Envelopes differs substantially from one device to another, and such a difference reduces as the devices' hardware stabilizes; i.e., the difference in this number across devices at minute 12 (Fig.~\ref{all-devices-12min}) is smaller than that observed at minute 1 (Fig.~\ref{all-devices-1min}). This trend has also been observed for the Q signal component, though not shown in the paper.

\section{The Effect of Hardware Stabilization on I/Q Data-Driven Device Fingerprinting}
\label{sec:effect}
We now turn our focus on explaining and demonstrating the effect of the warm-up time needed for the transceiver hardware to stabilize on the achievable performances of the widely used RF-driven device fingerprinting approaches that leverage deep learning models to extract device-specific features from raw I/Q data. Despite the rich amount of literature available on this topic, it is our understanding that the effect of hardware stabilization has not been carefully considered when RF data is collected and used to train and test such models. And it is our goal here to shed some light on what could go wrong had such stabilization aspects not been carefully accounted for when evaluating I/Q data-driven device fingerprinting approaches.
\subsection{Testbed Description and Dataset Collection Setup}
\label{sec:setup}
\begin{figure}
     \subfloat[15 Pycom Transmitters.\label{subfig-1:tx}]{%
       \includegraphics[width=0.233\textwidth, height = 0.237\textwidth]{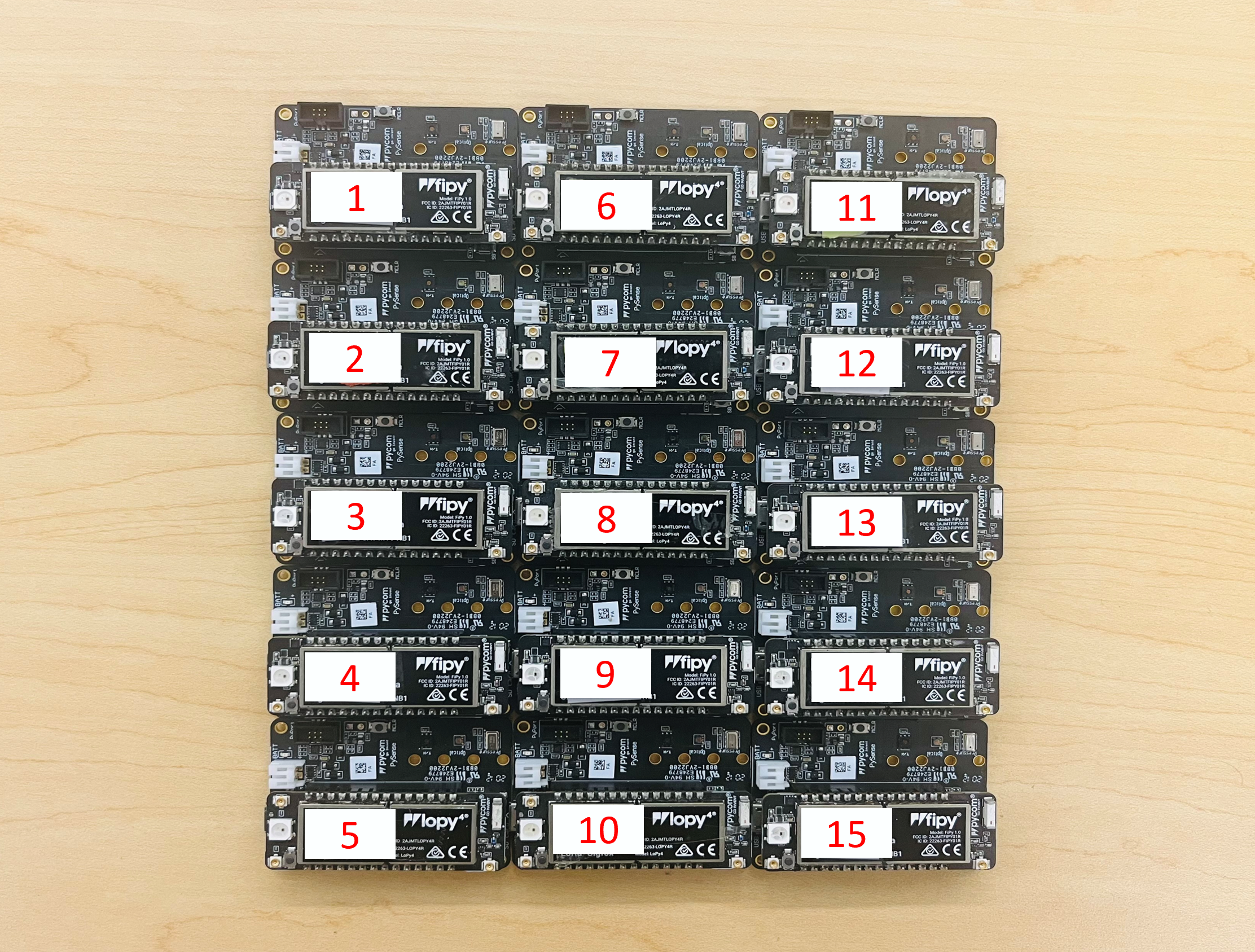}
     }
    \hspace{0.00001cm}
     \subfloat[Wired-WiFi and Wireless-WiFi.\label{subfig-2:recv}]{%
       \includegraphics[width=0.233\textwidth, height = 0.237\textwidth]{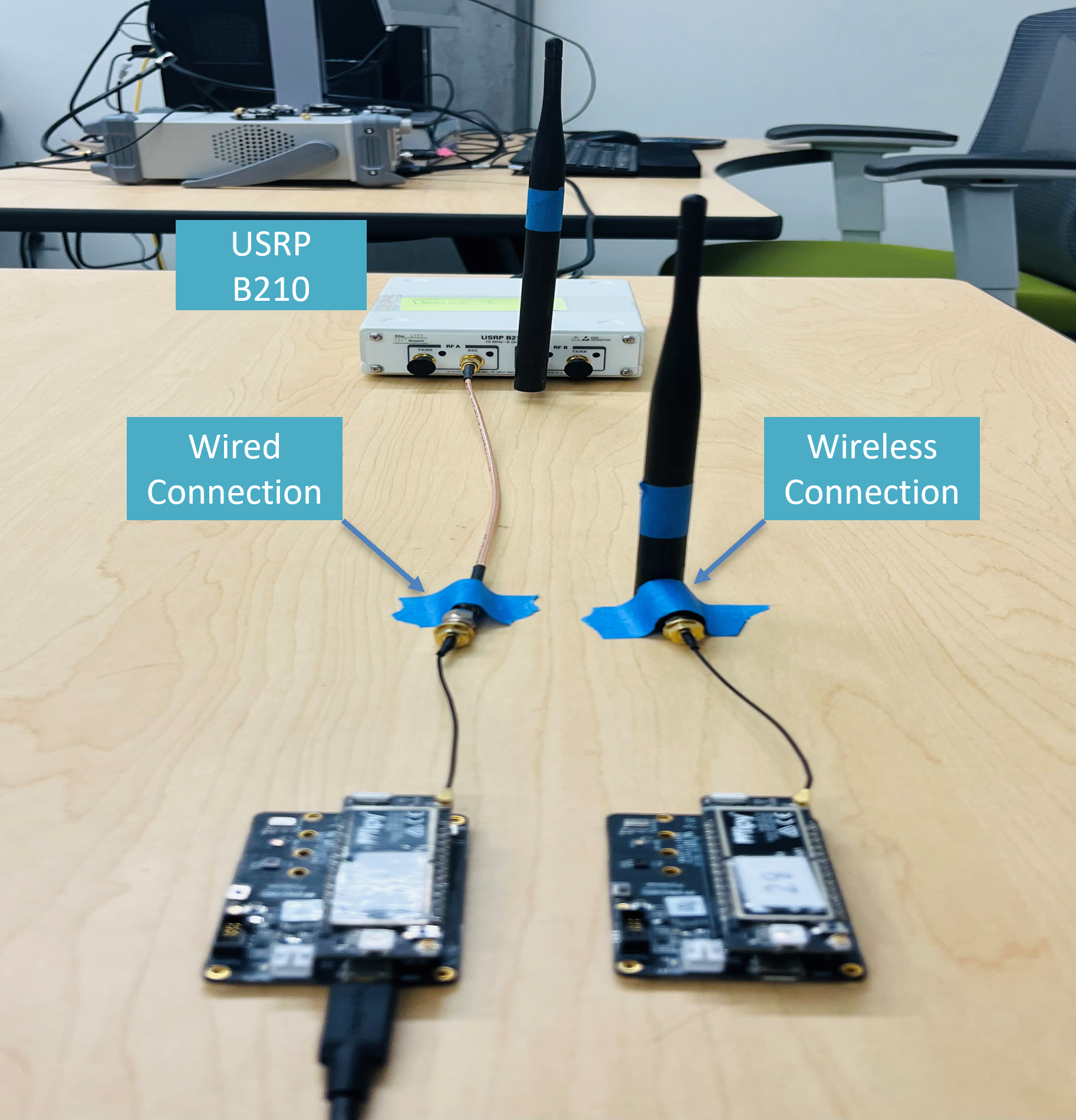}
     }
     \caption{IoT Testbed consisting of 15 Pycom transmitting devices and a USRP B210 receiving device}
     \label{fig:testbed}
\end{figure}
Our testbed is depicted in Fig.~\ref{fig:testbed} and comprises $15$ Pycom ($10$ FiPy and $5$ LoPy) devices. 
Data acquisition was performed using an Ettus USRP B$210$ receiver, which was synchronized with an external, oven-controlled crystal oscillator (OCXO) for improved sampling accuracy and stability. All devices were powered via USB from an HP laptop and configured to transmit IEEE802.11b WiFi packets using the high-rate direct-sequence spread-spectrum (HR/DSSS) physical-layer mode. All devices transmitted at a rate of 1Mbps with a carrier frequency of 2.412GHz and a bandwidth of 20MHz, while connected to the same 1/2 Wave Whip antenna.

\subsubsection{I/Q Data Collection}
We initiated the data collection process by powering on the Pycom devices one at a time, each configured to operate over WiFi Channel 1. The transmitters were programmed to transmit identical IEEE 802.11b WiFi frames with a duration of 559us back-to-back, separated by a small gap. We captured the first two minutes of transmissions using the USRP B210 at a sample rate of 45MSps. The captured signals were then digitally down-converted to the baseband and stored as I/Q samples on our computer. We refer to these initial captures as ``unstable capture'' since the crystal oscillators on the Pycom devices had not yet stabilized during this time. After allowing the devices sufficient time to stabilize, we continued with the data collection process. We waited an additional 10 minutes before initiating the capture process again. This ensured that the crystal oscillators had stabilized and were operating consistently. We refer to these subsequent captures as ``stable captures''. To avoid any data dependency about the identity of the WiFi transmitter, all devices were configured to broadcast the same packets, which include the same spoofed MAC address and a payload of zero-bytes.
Finally, we extracted the WiFi packets from the raw I/Q sample files and stored them in HDF5-formatted files in the same order they were received. This method allowed us to maintain the integrity of the captured signals and ensured that they were accurately represented in the final dataset. The created datasets can be downloaded from NetSTAR Lab at 
\href{https://research.engr.oregonstate.edu/hamdaoui/datasets/}{\color{blue}{http://research.engr.oregonstate.edu/hamdaoui/datasets}}.

\subsubsection{Experimental Setups} \label{dataset_Section}
Our datasets of WiFi data, collected from $15$ Pycom devices, are captured using both wireless and wired setups, each over three consecutive days. For each scenario, we captured data both during and after device stabilization.
\begin{itemize}
\item {\bf Wireless Setup:} \label{wireless_Setup}
All transmitters were placed $1$m away from the USRP receiver, equipped with a VERT$900$ antenna. We repeated this experiment over three days to study the robustness of the fingerprinting over time. 
This generated $5000$+ WiFi frames per device every day. 

\item {\bf Wired Setup:} \label{wired_Setup}
To factor out the wireless channel's impact, we wired the transmitters to the USRP receiver using an SMA cable and repeated this experiment over three days. This generated $3000$+ WiFi frames per device every day.  

\end{itemize}

\subsection{Device Fingerprinting Evaluation}
\label{sec:results}

For evaluation purposes, we used the CNN (convolution neural network) model employed in~\cite{elmaghbub2021} as the deep learning classifier, and considered the following three experiments: 

\begin{itemize}
\item \textbf{Experiment 1: Train on min12-captures and Test on min1-captures.} 
We train the CNN model using WiFi packets collected after hardware stabilization (after 12 minutes from device activation), and test it using packets captured right after (within the first two minutes) device activation on the same or on a different day. 
This experiment allows the evaluation of the impact of the warm-up period on the fingerprinting accuracy. 

\item \textbf{Experiment 2: Train on min12-captures and Test on min12-captures.} 
We train the model using WiFi packets collected after hardware stabilization and test it also using stabilized WiFi captures collected on the same or on a different day.
This allows the evaluation when training and testing on data captured after device stabilization. 

\item \textbf{Experiment 3: Train on min1-captures and Test on min1-captures.} 
We train the CNN model using packets captured right after device activation and test it using packets obtained, also right after device activation, on the same or on a different day. This enables to assess the underlying behavior of the I/Q signal and fingerprinting performance at the beginning of the warm-up period.
\end{itemize}

We adopt 5-fold cross-validation where the data is divided into five equally-sized, non-overlapping partitions, with four partitions (comprising 3200 packets) utilized for training and the firth partition (consisting of 800 packets) utilized for testing. To represent each packet, we employ a 2x8192 tensor, encompassing 8192 samples of each of the I and Q data. This window size is selected experimentally based on performance superiority.

\subsubsection{Experiment 1 Results: Classification Accuracy When Training on min12-captures and Testing on min1-captures}
\begin{figure}[t!]

   \subfloat[Accuracy (Wireless)]{
   \includegraphics[width=.47\columnwidth]{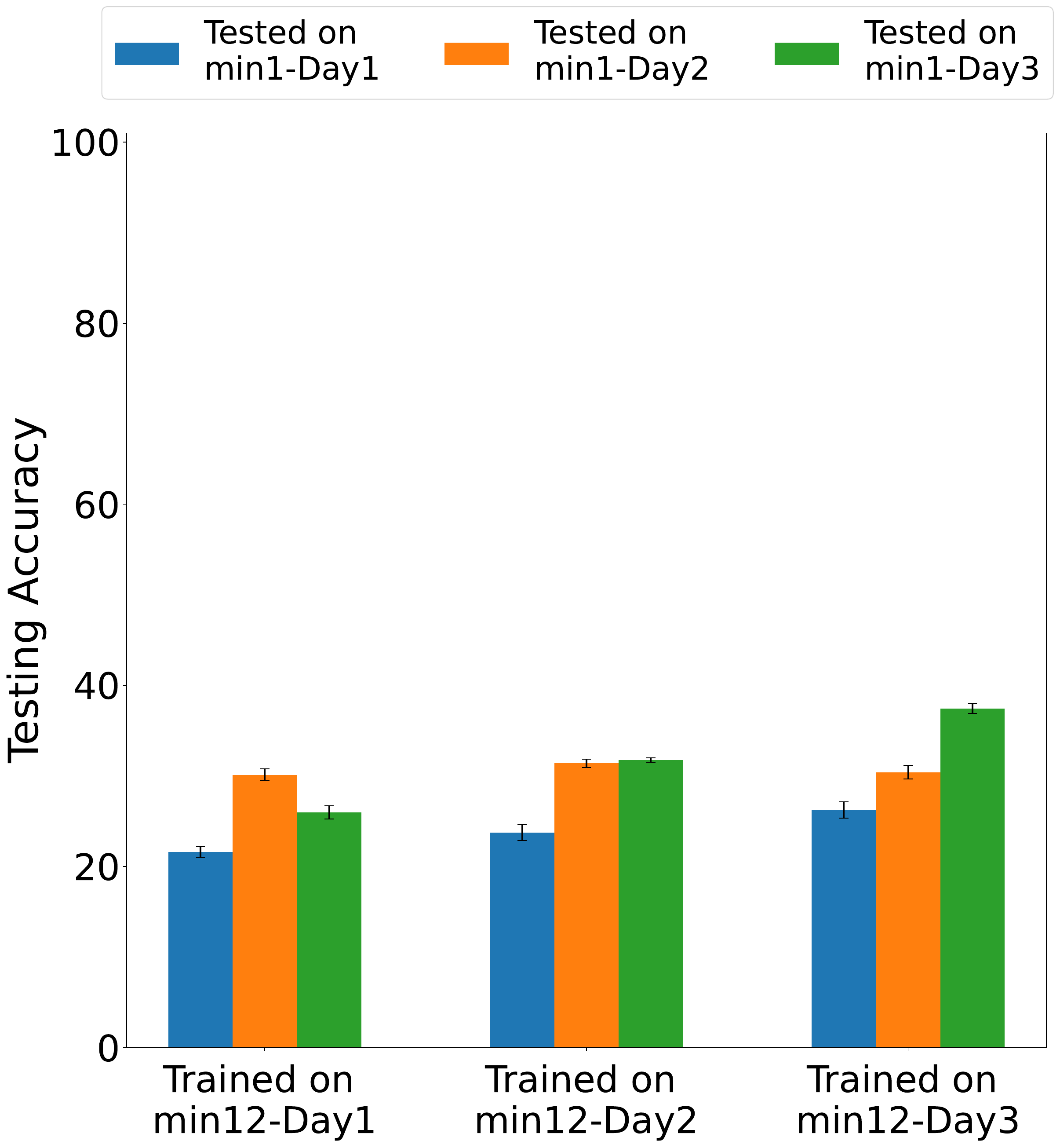}
   \label{cap1-12-wireless}}
   \hfill
   \subfloat[Accuracy (Wired)]{
   \includegraphics[width=.47\columnwidth]{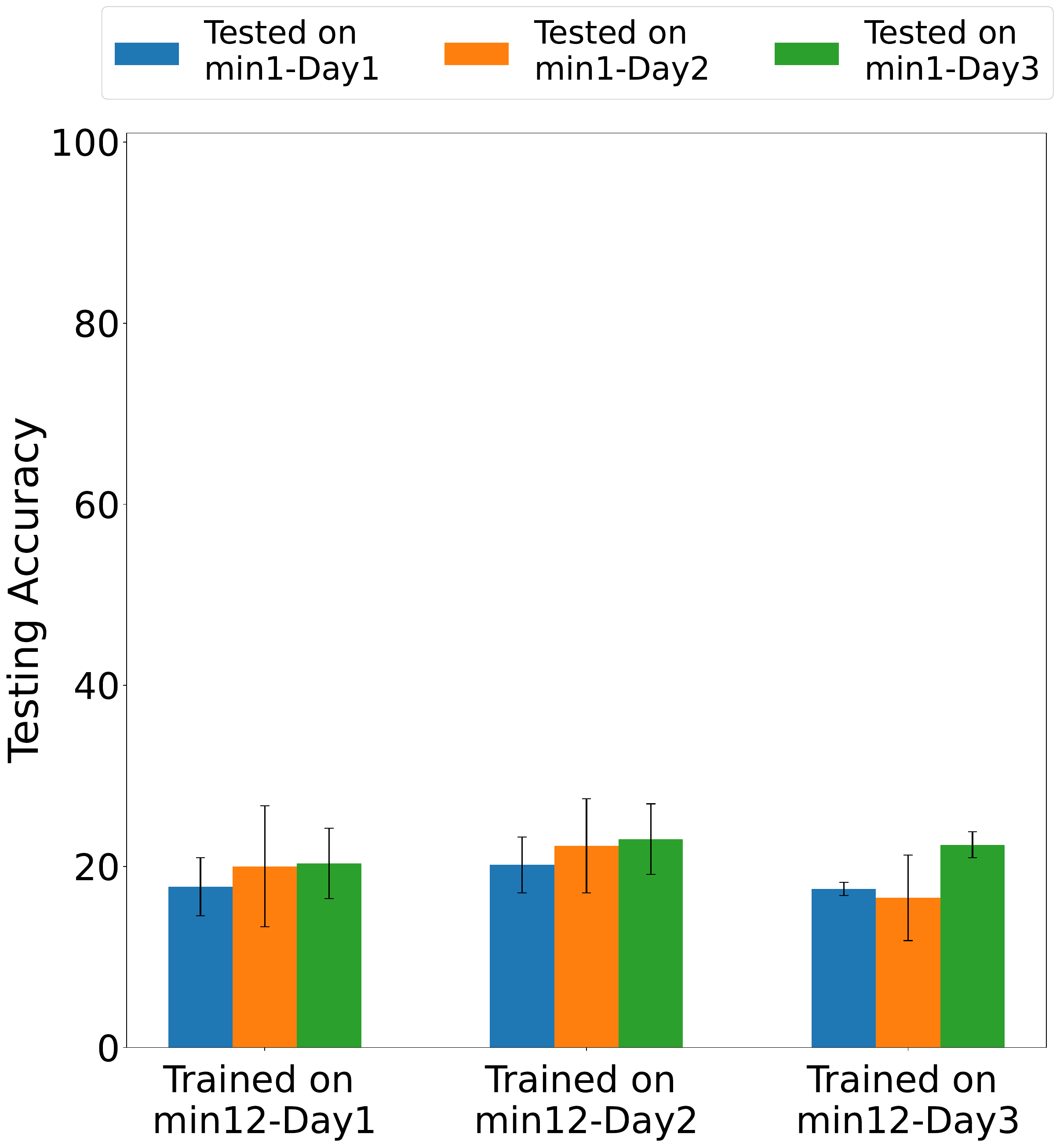}
   \label{cap1-12-wired}} 
   \\
   \subfloat[Confusion Matrix (Wireless)]{
   \includegraphics[width=.47\columnwidth]{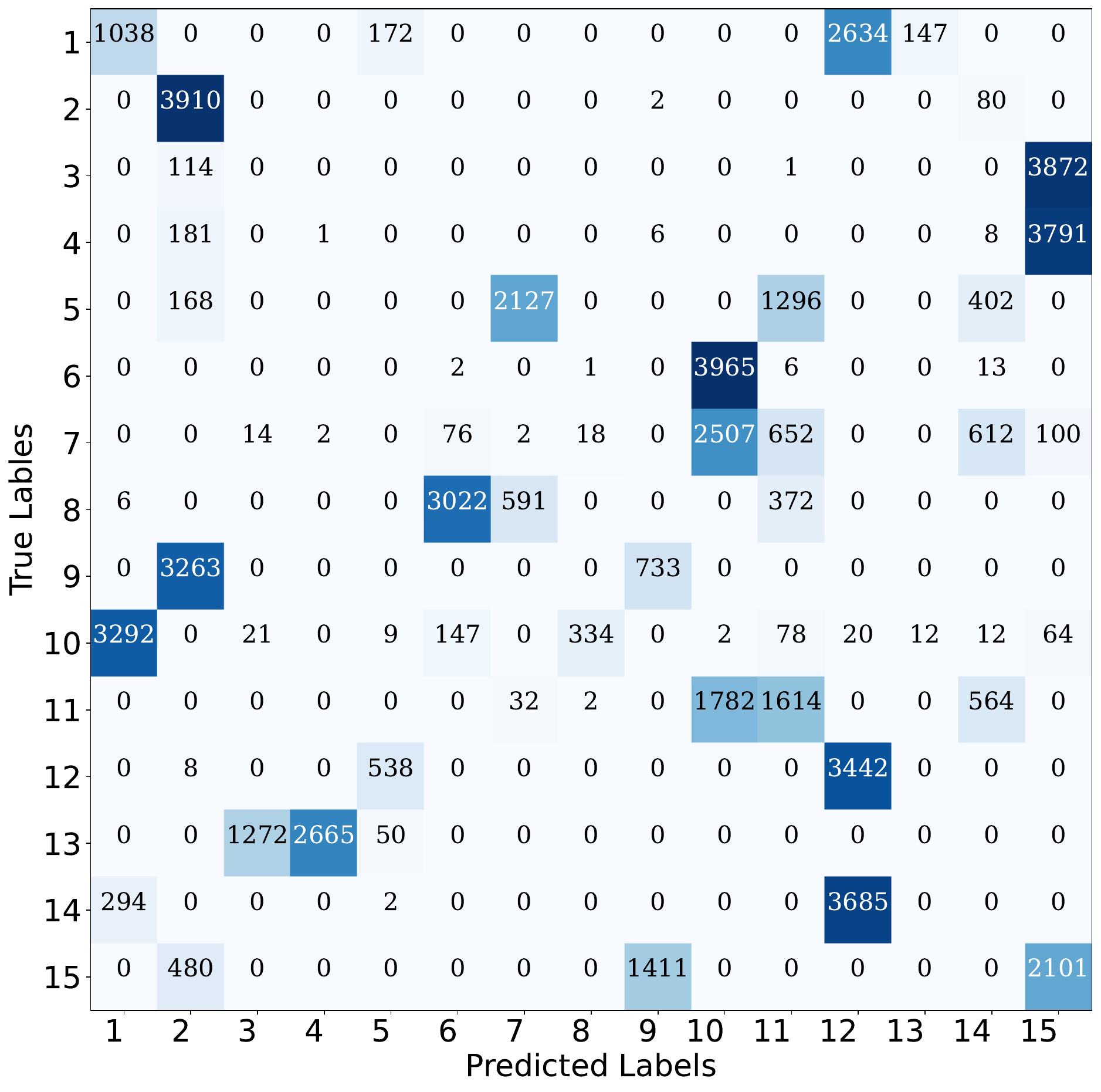}
   \label{cap1-2-day3}}
   \hfill
   \subfloat[Confusion Matrix (Wired)]{
   \includegraphics[width=.47\columnwidth]{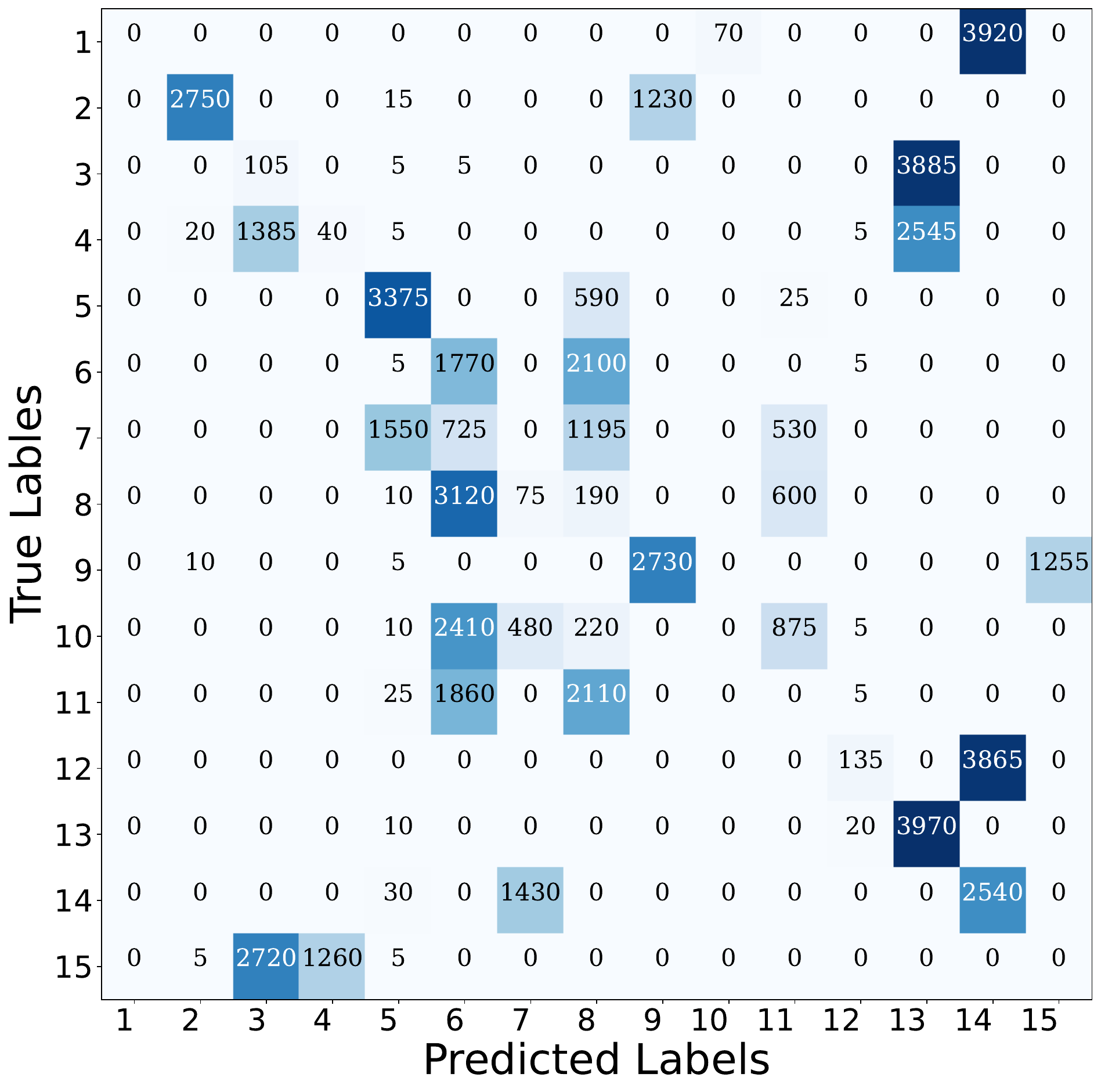}
   \label{cap1-12-day2}}
   \caption{Experiment 1: training on min12-captures and testing on min1-captures. Confusion matrices are when training on min12-captures (day 1) and testing on min1-captures (day 1).}
\label{IQ-days}
\end{figure}
The obtained results for both the wireless (Fig.~\ref{cap1-12-wireless}) and wired (Fig.~\ref{cap1-12-wired}) settings provide compelling evidence of the sensitivity of the RF fingerprinting classifier to the hardware warm-up and stabilization. 
For both scenarios, the environment remained static and the time gap between training and testing is as short as 12 minutes. 
Remarkably, these results show that the average classification accuracy degrades significantly and drops to as low as 17.8\% when the model is trained on data captured at minute 12 (min12-captures), but tested on data captured at minute 1 (min1-captures) after device activation. 
The performance worsens when considering cross-day scenarios. For example, for the case of the wireless setting shown in Fig.~\ref{cap1-12-wireless}, when training is done using min12-captures of day 1, the classification accuracies are 21.6\%, 30.4\%, and 26.7\% when testing is done using min1-captures of days 1, 2, and 3, respectively. Similar trends are observed when the model is trained on captures from days 2 or 3 and tested on the other-day captures, as well as whether the wired or wireless setting is used. The corresponding confusion matrices given in Fig.~\ref{cap1-2-day3} and Fig.~\ref{cap1-12-day2} visually portray the challenges faced by the learning models in adapting to the evolving signal characteristics during the 12-minute warm-up period.
The consistency of these observed trends across both the wired and wireless settings questions the prevailing belief that attributes the time-sensitivity of deep learning-based RF fingerprinting methods solely to the wireless channel impact. Our findings confirm that the signal characteristics change significantly during hardware warm-up time, which result in drastic degradation of the RF fingerprinting performance.

\subsubsection{Experiment 2 Results: Classification Accuracy When Both Training and Testing are on min12-captures}
\begin{figure}[t!]
  
\subfloat[Accuracy (Wireless)]{
   \includegraphics[width=.47\columnwidth]{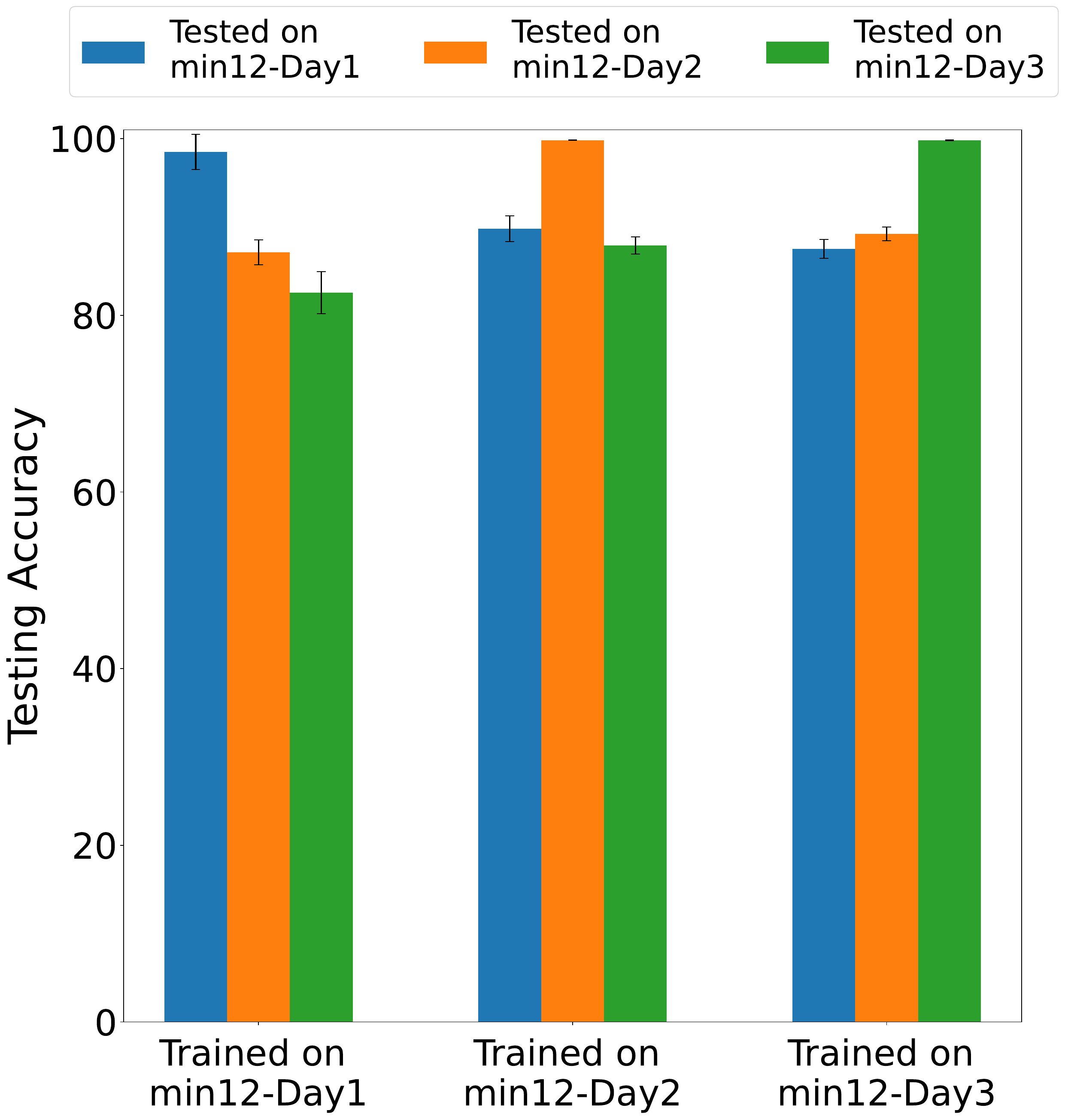}
   \label{IQ-trained-days}}
    \hfill
   \subfloat[Accuracy (Wired)]{
   \includegraphics[width=.47\columnwidth,]{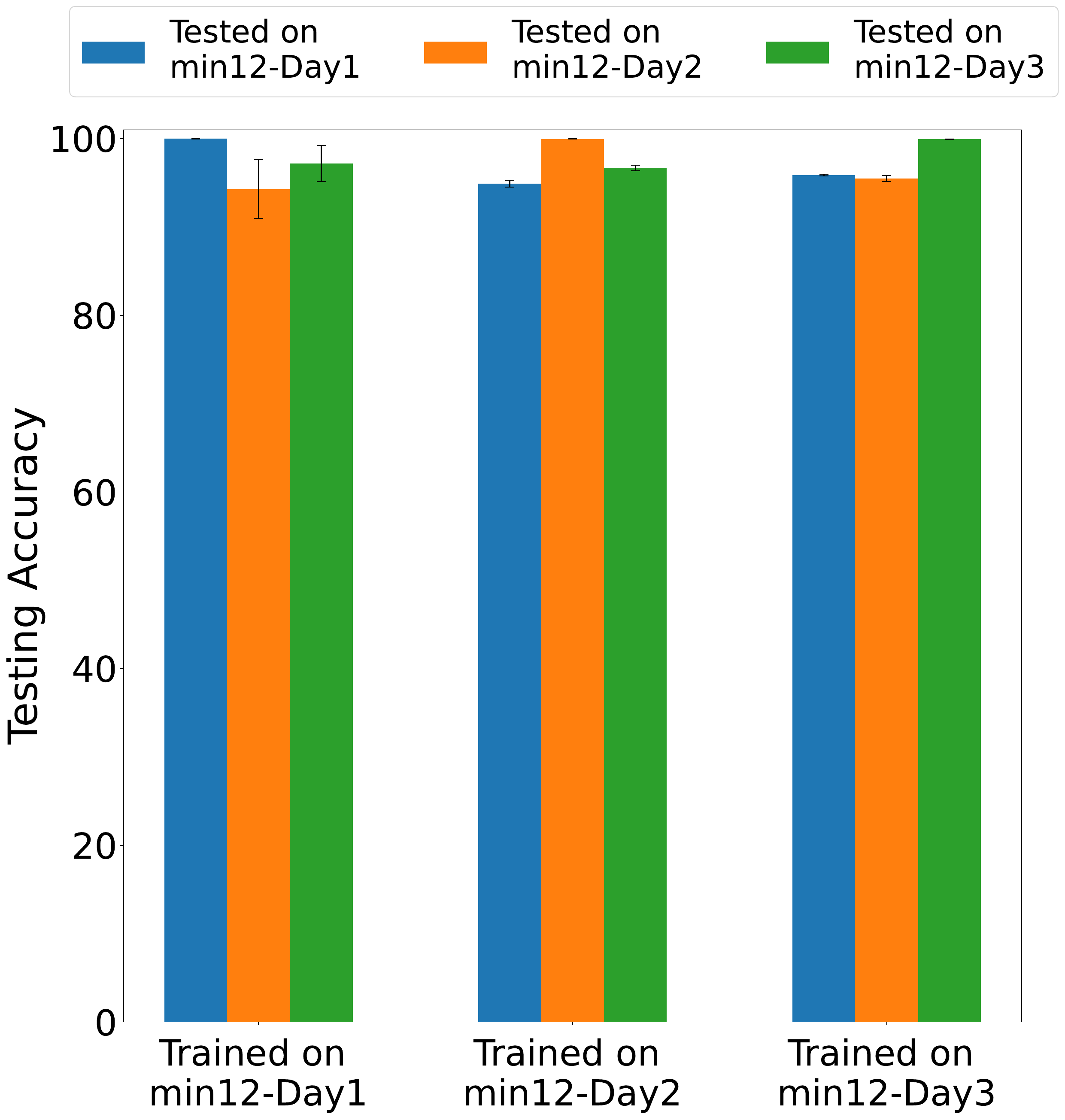}
   \label{stable-wired}}
    \\
\subfloat[Confusion Matrix (Wireless)]{
   \includegraphics[width=.47\columnwidth]{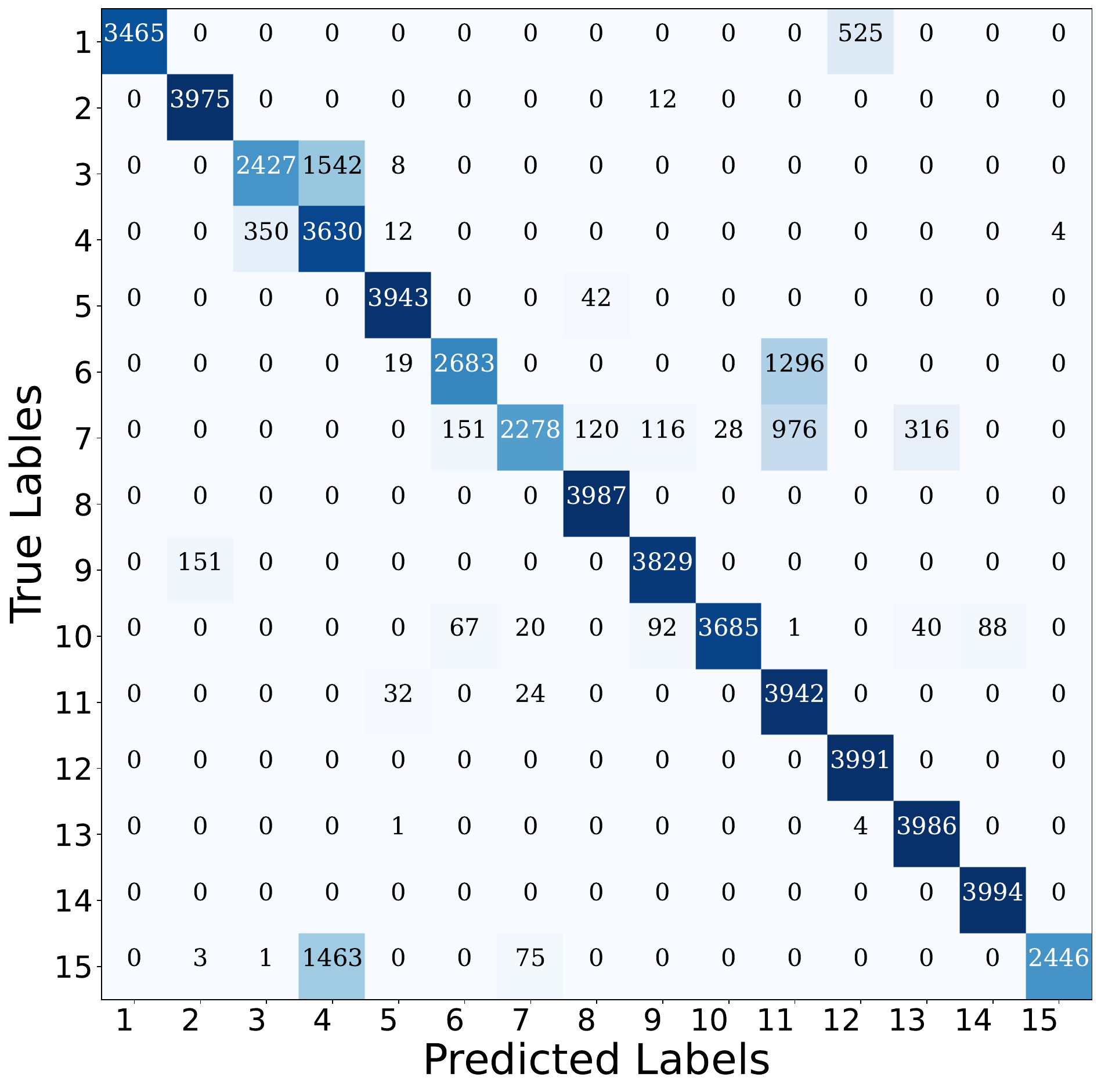}
   \label{IQ-day2}}
   \hfill
   \subfloat[Confusion Matrix (Wired)]{
   \includegraphics[width=.47\columnwidth]{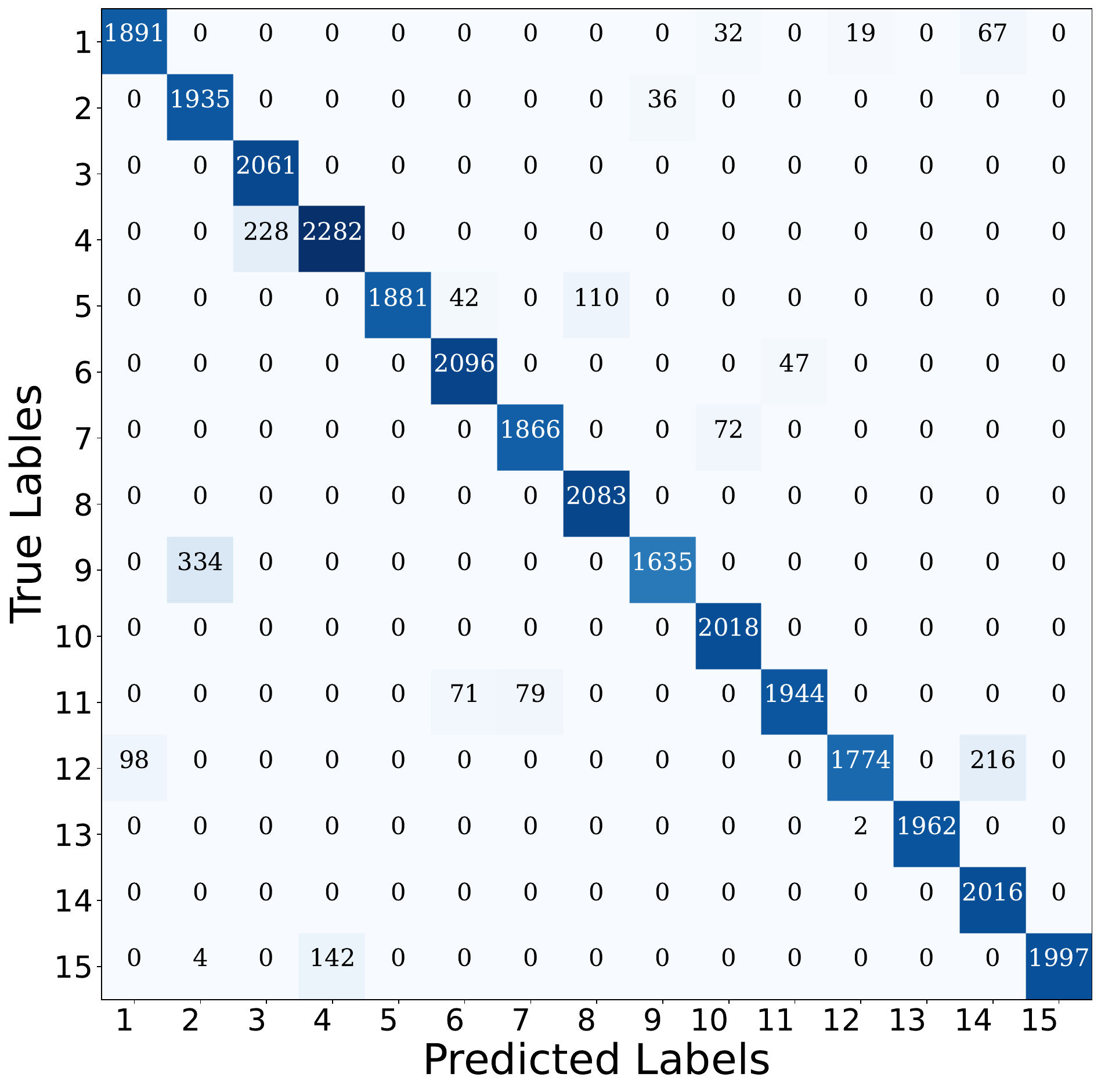}
   \label{IQ-day3}}
   \caption{Experiment 2: training and testing on min12-captures. Confusion matrices are when training on min12-captures (day 1) and testing on min12-captures (day 2).}
\label{IQ-days}
\end{figure}

Fig. \ref{IQ-trained-days} illustrates the obtained accuracy results for the wireless setting, when training and testing are both done on data collected after device hardware stabilization, which is experimentally shown to be about 12 minutes in our case. 
Observe the high accuracy (nearly 99\%) that the models are able to achieve when training and testing are done using same-day min12-captures. Remarkably, the model also demonstrates strong performance in cross-day classification. For instance, when trained on day 1's min12-captures and tested on min12-captures from day 2 or day 3, the achieved cross-day testing accuracies of 87.3\% or 82.3\%, respectively, are still high compared to what was observed in Experiment 1 above when data is collected during warm-up time.
Similarly, when the model is trained on day 2's min12-captures and tested on min12-captures of day 1 or day 3, the average cross-day testing accuracy is 89.8\% or 87.9\%, respectively. Moreover, when trained on day 3 and tested on days 1 or 2, the testing accuracy is 87.5\% or 89.2\%, respectively. The aggregate confusion matrix of the cross-day testing is provided in Fig. \ref{IQ-day2}. 

In the wired setting, the model performs even better (see Fig.~\ref{stable-wired}), achieving cross-day testing accuracy of 94.9\% (resp. 95.9\%) when trained on day 1's min12-captures and tested on day 2's (resp. day 3's) min12-captures. Notably, the average performance drop across the three days, when trained on one day and tested on one of the other two days, is 4.4\%. In contrast, the performance drop in the wireless setting is higher, at 12\%. This 8\% difference underscores the influence of the channel in a static environment. On the other hand, the substantial disparity in performance drop observed when the training packets are captured before and after hardware stabilization highlights the profound impact of hardware stability on the performance of RF fingerprinting methods.

\subsubsection{Experiment 3 Results: Classification Accuracy When Both Training and Testing are on min1-captures}

\begin{figure}[t!]
\subfloat[Testing Accuracy]{
   \includegraphics[width=.47\columnwidth, height=.47\columnwidth]{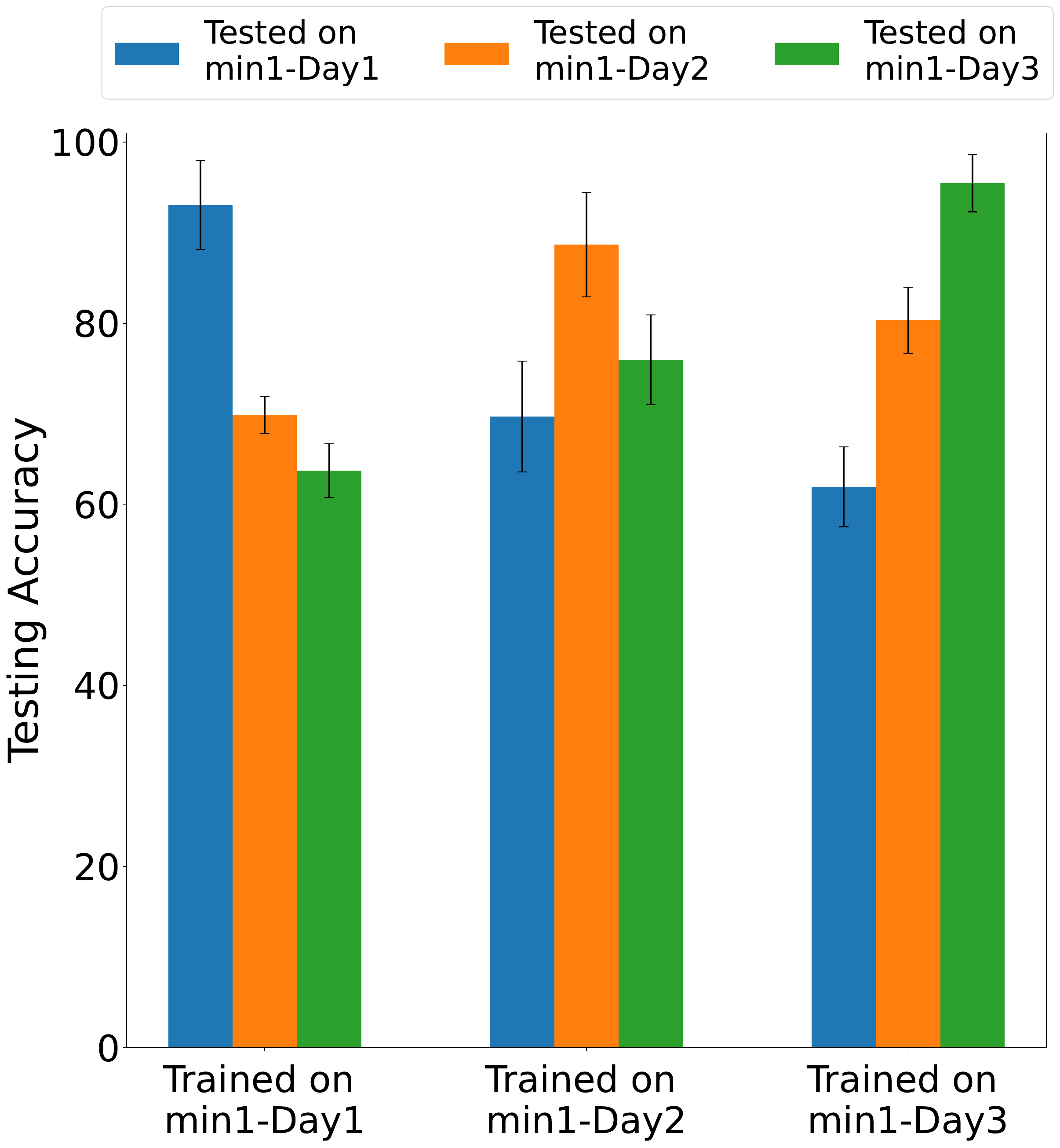}
   \label{cap1_days}}
   \hfill
\subfloat[Confusion Matrix]{
   \includegraphics[width=.47\columnwidth]{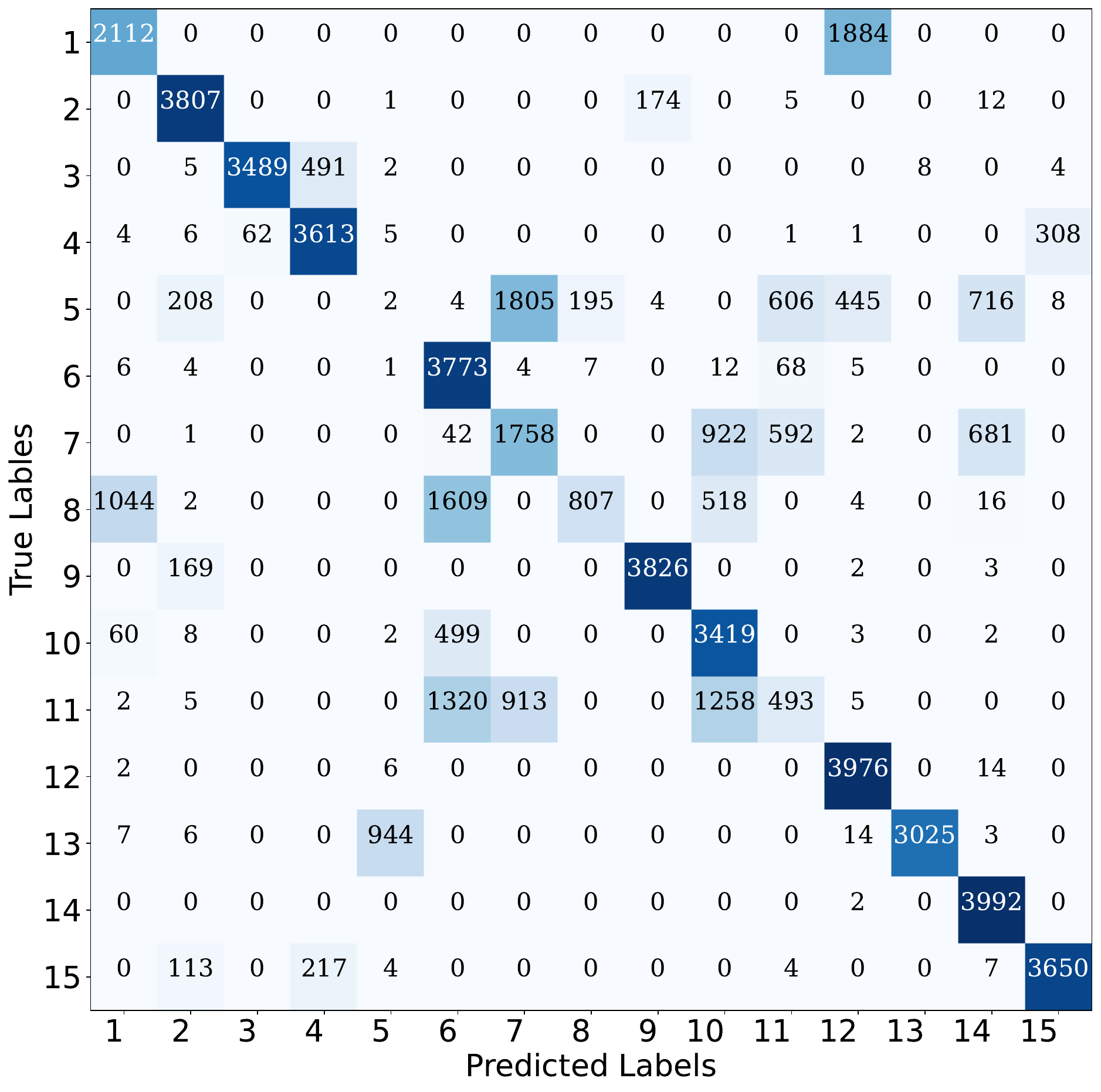}
   \label{cap1-1_1}}\\

   \caption{Experiment 3: training and testing on min1-captures on the wireless scenario. Confusion matrices are when training on min1-captures (day 1) and testing on min1-captures (day 2).}
\label{exp2}
\end{figure}

We now study the behavior of the device fingerprinting accuracy when training and testing are done at the beginning of the warm-up period and across different days. For this, we train the model on min1-captures (data captured within the first two minutes) of one day and tested it also on min1-captures but from a different day. In this experiment, we only consider the wireless setting. The findings presented in Fig. \ref{cap1_days} demonstrate an average testing accuracy of 70\% or 62\% when the model is trained on min1-captures of day 1 and tested on min1-captures of day 2 or day 3, respectively, which is considerably higher than the achieved accuracy when training on min12-captures of the same day, i.e., 21.6\%. These findings indicate a systematic drift in signal characteristics during the stabilization period, with consistent behavior observed across days. Therefore, further investigations are needed to gain a comprehensive understanding of this behavior and exploit it for improving RF fingerprinting methods. Additionally, the confusion matrix presented in Fig. \ref{cap1-1_1} again highlights the challenges faced by the model in recognizing particular devices during the warm-up phase, particularly for Devices 5, 7, 8, and 11, indicating that different devices can be more stable than others during the warm-up period. These results do underscore the importance of considering hardware stabilization when developing RF fingerprinting techniques.

\section{The Cause Behind the Observed Behavior of the I/Q Signals' Envelopes}
\label{sec:inaccuracy}
Our experimental studies of the impact of the device hardware warm-up \& stabilization, presented in Section~\ref{sec:stability}, reveal two important trends regarding the behavior of the Envelopes of the received I/Q signals. Recall again that the Envelope of the oscillating I/Q signal is the smooth boundary that outlines the extremes of the signal~\cite{johnson2011software}.
\begin{enumerate}
    \item For a given device, the number of 'humps' in the I/Q signal's Envelope increases with time during hardware warm-up and eventually settles to a `stabilized shape' after some time from device activation. The Pycom transmitters that we tested in our experiments stabilized after about 12 minutes.
    
    \item The number humps in the envelopes differs from one device to another, and the difference across devices exists right when the devices are powered on and continues to exist until after the devices are stabilized; i.e., after about 12 minutes from device activation.
\end{enumerate}
The question that arises then is: what is the cause of such a behavior of the I/Q signals? Our research shows that the main cause is the instability and inaccuracy of the local oscillator's frequency, which varies across devices due to the oscillator hardware imperfections/impairments incurred during manufacturing and varies over time due to the hardware instability during warm-up. In this section, we confirm and validate this hypothesis through both simulation and experimentation.

\subsection{The Carrier Frequency Offset (CFO)}
Local oscillators are an essential hardware component in the RF transceiver chain with the main function of producing oscillating signals that are used for signal up-conversion (at sender) and down-conversion (at receiver). The inaccuracy and instability of the frequency of such oscillating signals, typically caused by external factors like temperature, vibration and electromagnetic interference, impact various aspects of the overall system performance behavior. 
Various types of crystal oscillators have been developed over the years to improve their robustness to these external factors, including temperature-controlled crystal oscillators (TCXOs), which feature temperature compensation, and oven-controlled crystal oscillators (OCXOs), which place the crystal in a temperature-controlled environment to maintain a constant temperature \cite{zhou2008frequency}.

One important aspect resulting from the oscillating frequency inaccuracy is the inevitable mismatch between the receiver's local oscillating frequency and that of the sender. This mismatch in frequency is known as the {\em Carrier Frequency Offset or CFO} and often leads to signal distortion and reduced communication quality. CFO has been leveraged to serve as RF device fingerprints that differentiate between WiFi devices (e.g., \cite{vo2016fingerprinting, CFO_clockSyn}).
In this work, we show that this CFO impairment is the main cause behind the observed behavior of the I/Q signals' envelopes. 
In other words, we show that CFO values do change over time during hardware stabilization (due to hardware warm-up) and differ across devices (due to manufacturing imperfection of the oscillators).
From a device fingerprinting viewpoint, it is worth mentioning, though, that while the change that exists across devices can be viewed as a blessing, as it could be leveraged to provide distinguishing device fingerprints, the change over time during hardware warm-up could result in fingerprint confusion and hence could be viewed as a curse for device fingerprinting.
Therefore, without carefully considering their stability and variability during hardware warm-up, CFO impairments may not serve as robust and consistent fingerprints that can be relied on to distinguish among devices.

Next, we show and confirm that the CFO is what is behind the I/Q envelope behavior observed and discussed in Section~\ref{sec:stability}.


\subsection{CFO Impact Analysis Through Matlab Simulation}
\begin{figure}
\centering
    \begin{minipage}{\linewidth} 
    \centering 
    \subfloat[CFO = 0 (ideal scenario)]{
    \includegraphics[width=.45\linewidth,height=0.42\linewidth]{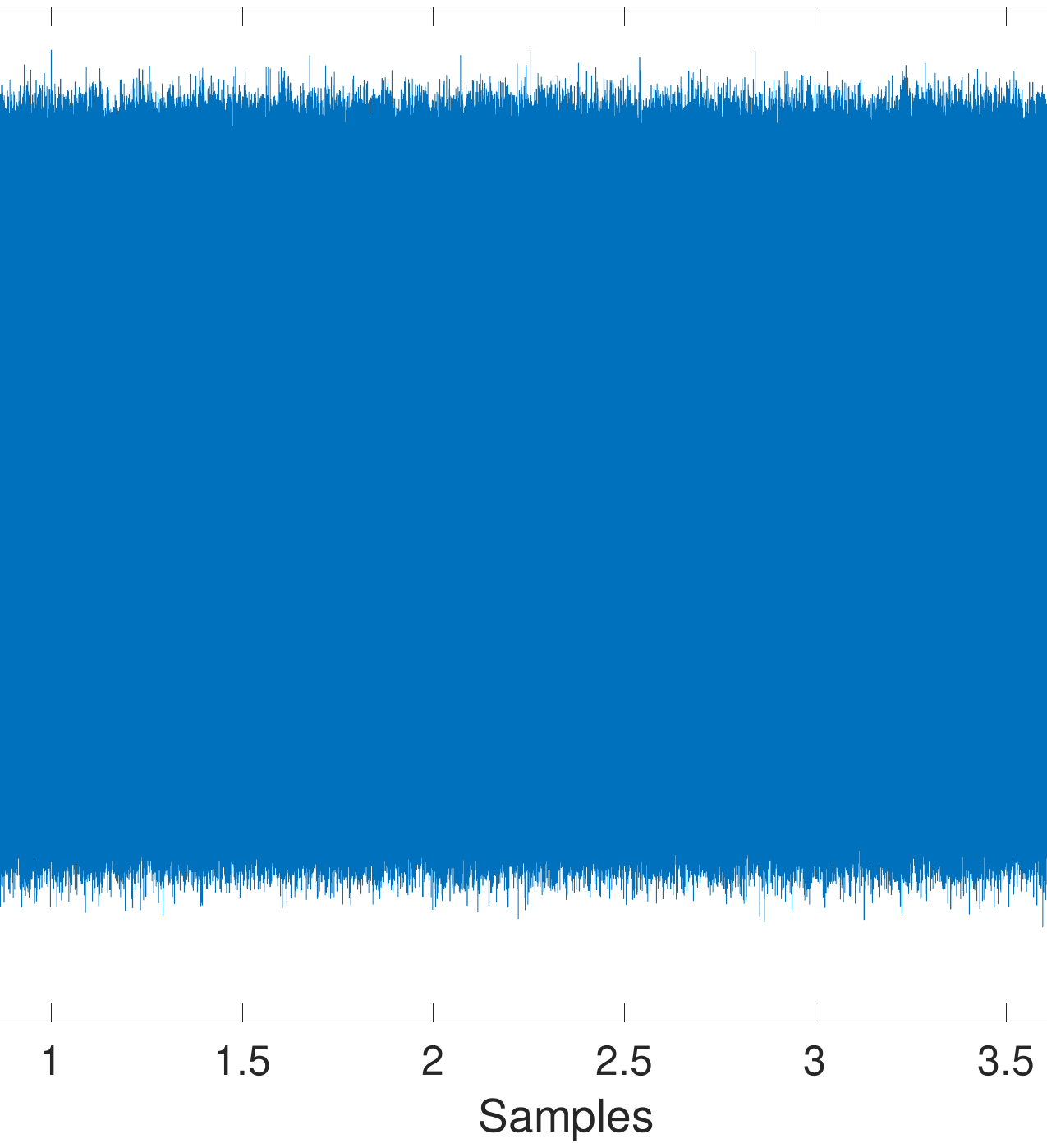} 
     \label{cfo0}}
     \hfill
    \subfloat[CFO = 50 Hz]{
    \includegraphics[width=.45\linewidth,height=0.42\linewidth]{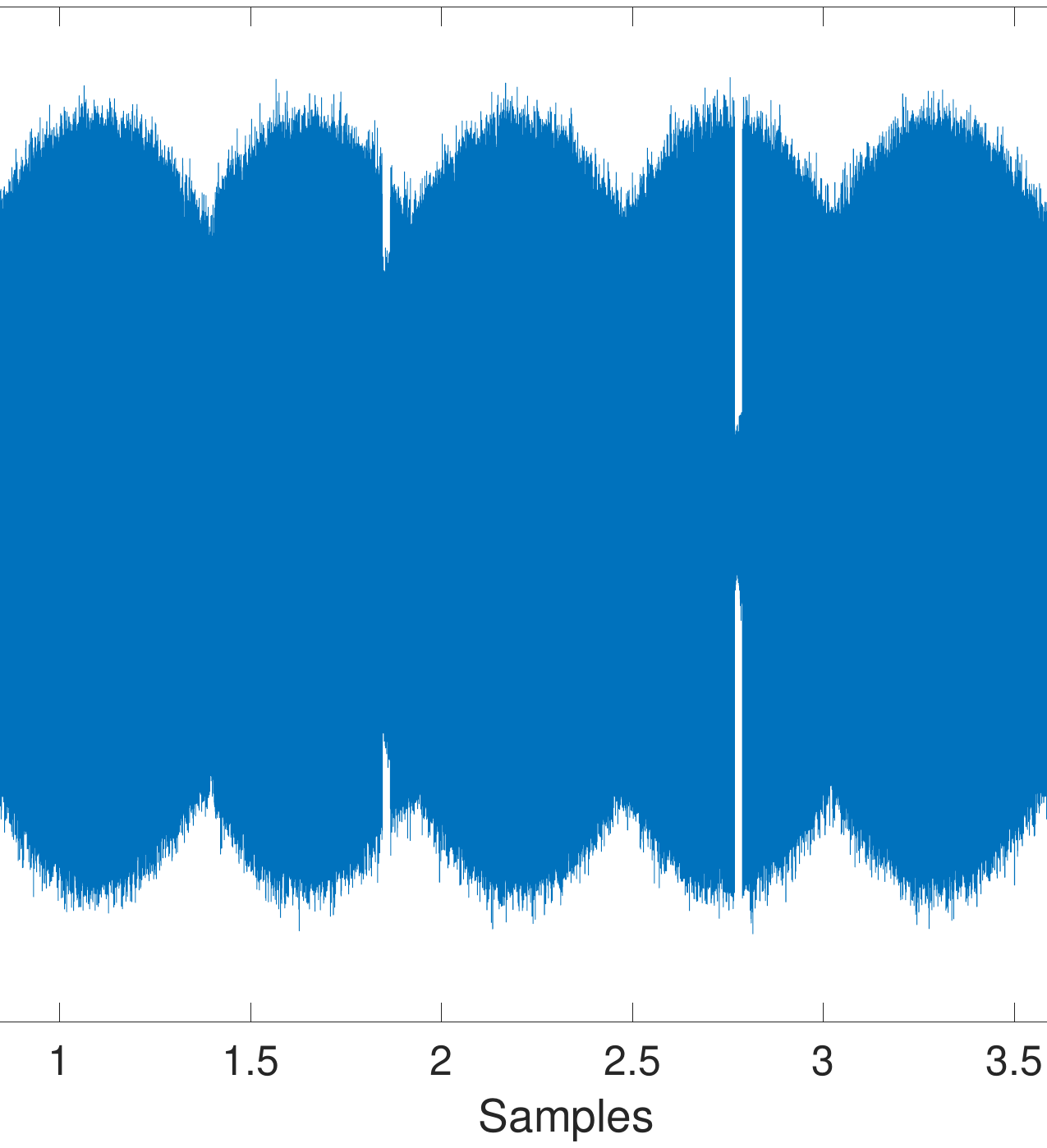}
    \label{cfo50}}\\
    \subfloat[CFO = 100 Hz]{
    \includegraphics[width=.45\linewidth,height=0.42\linewidth]{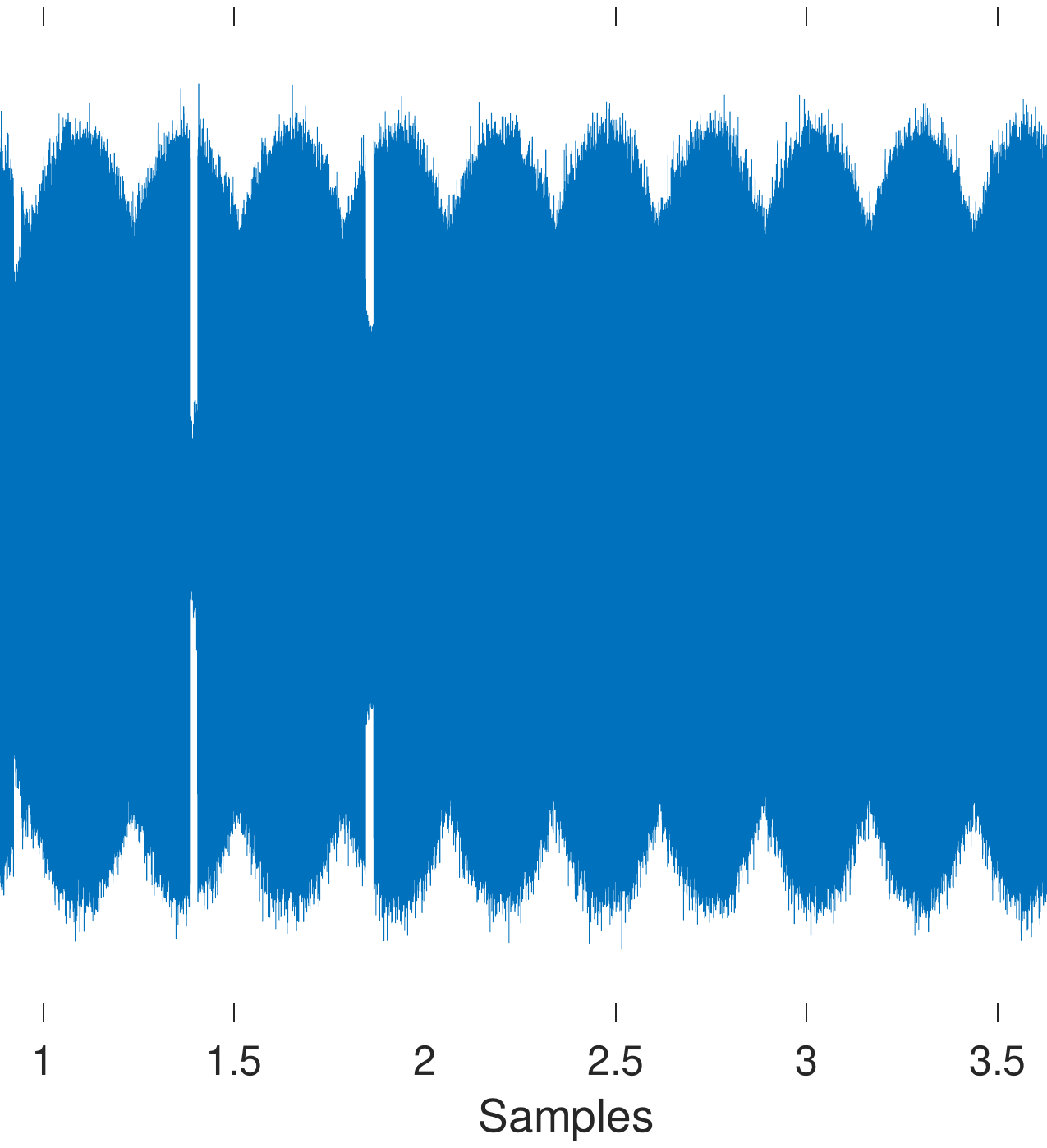}
    \label{cfo100}}
     \hfill
\subfloat[CFO = 200 Hz]{
    \includegraphics[width=.45\linewidth,height=0.42\linewidth]{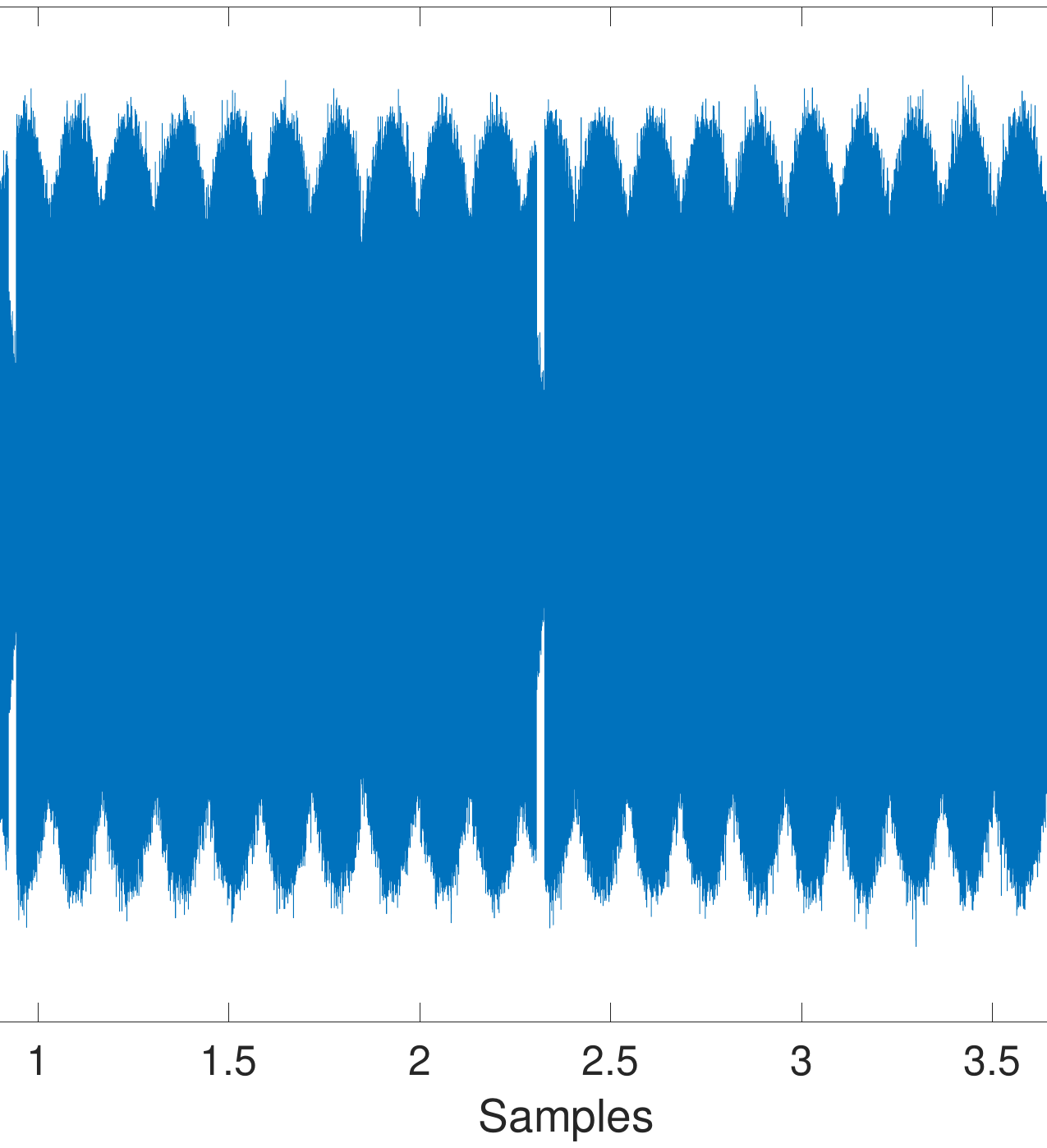}
    \label{cfo200}}
    \caption{Simulated time-domain I signal component. Y-axis is the amplitude of the I values; X-axis is the time in samples.} 
    \label{cfo-matlab} 
    \end{minipage}
\end{figure}

We used MATLAB R2023b to build our wireless communication system model and create I/Q datasets while varying the CFO value between the sending and the receiving devices. We used
MATLAB's WLAN toolbox to generate multiple IEEE 802.11b WiFi DSSS waveforms impaired with the following CFO values: 0 Hz (ideal device), 50 Hz, 100 Hz, and 200 Hz.
The CFO-impaired transmitted signal is then first passed through an AWGN channel, and then down-converted and sampled by the receiver to generate I/Q samples.
For each case, we collected $10$ WiFi frames, with each frame having a size of $1000$ bits. 
Then, we extracted the real (I) components of the signals and plotted them separately: Fig.~\ref{cfo0} for CFO = 0; Fig.~\ref{cfo50} for CFO = 50Hz; Fig.~\ref{cfo100} for CFO = 100Hz; and Fig.~\ref{cfo200} for CFO = 200Hz.
The simulated results clearly show the dependency between the CFO values and the number of humps in the envelopes of the time-domain I component of the I/Q signal, and that the CFO is what causes the observed envelope behavior. 
Note that for the ideal scenario when CFO = 0, the I signal exhibits a constant envelope, but when the CFO value in nonzero,  the signal presents a sinusoidal envelope whose number of 'humps' increases with the CFO value. 
The same trends were observed for the Q components as well, but we did not include them here.

We want to mention that we also varied other hardware impairments, including IQ imbalance, Phase Noise, and DC offset, and monitored the I/Q signal shape, but have not noticed any sinusoidal behavior of the signals' envelopes, thereby confirming that other impairments do not result in the same envelope behavior that we observed when varying the CFO.

\subsection{CFO Impact Analysis Through Modulation/Demodulation}
\begin{figure}
\centering
    \begin{minipage}[t]{\linewidth} 
    \centering 
    \subfloat[CFO = 0]{
    \includegraphics[width=\linewidth,height=0.22\linewidth]{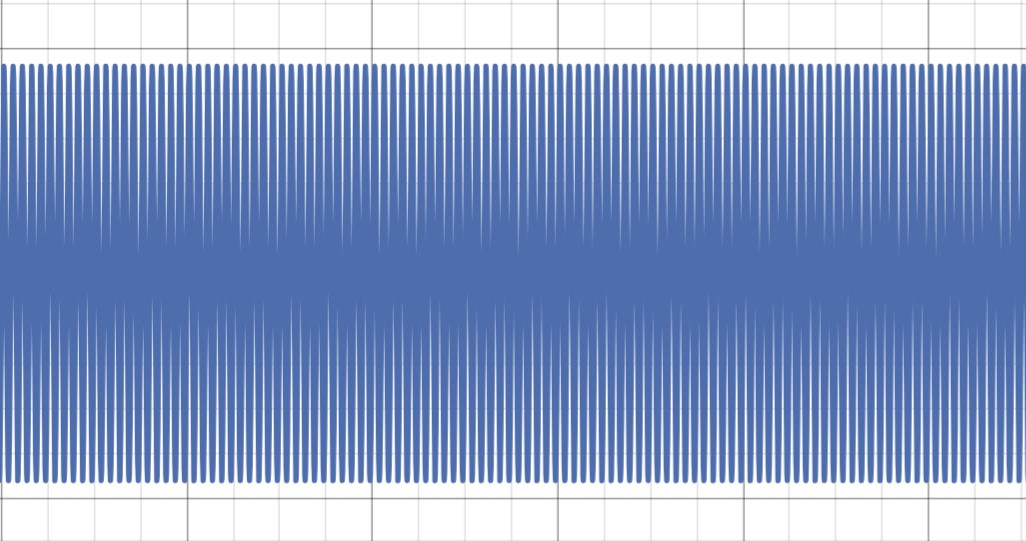} 
     \label{cfo-a0}}
     \hfill
    \subfloat[CFO = 0.1Hz]{
    \includegraphics[width=\linewidth,height=0.22\linewidth]{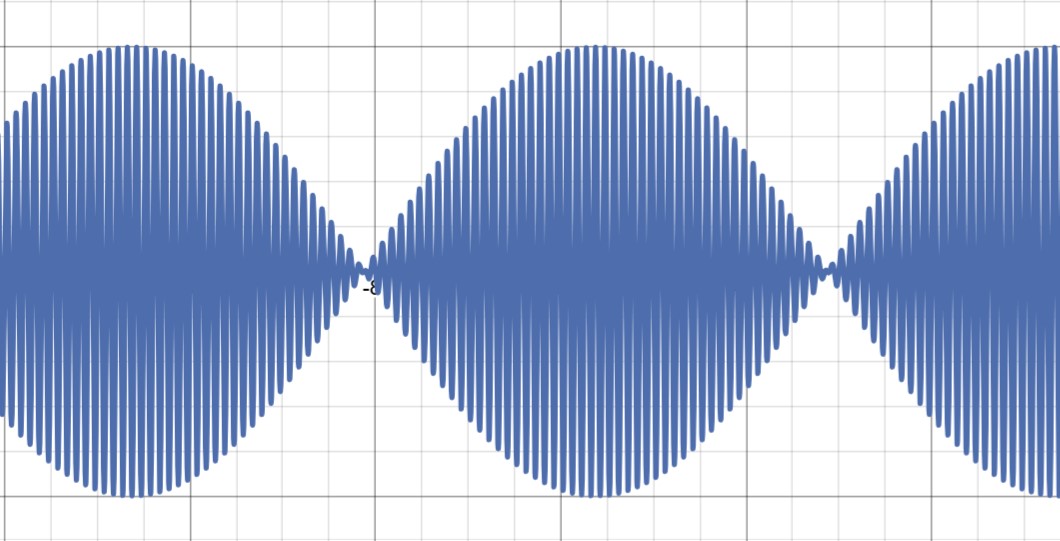}
    \label{cfo-a01}}\\
    \subfloat[CFO = 0.2Hz]{
    \includegraphics[width=\linewidth,height=0.22\linewidth]{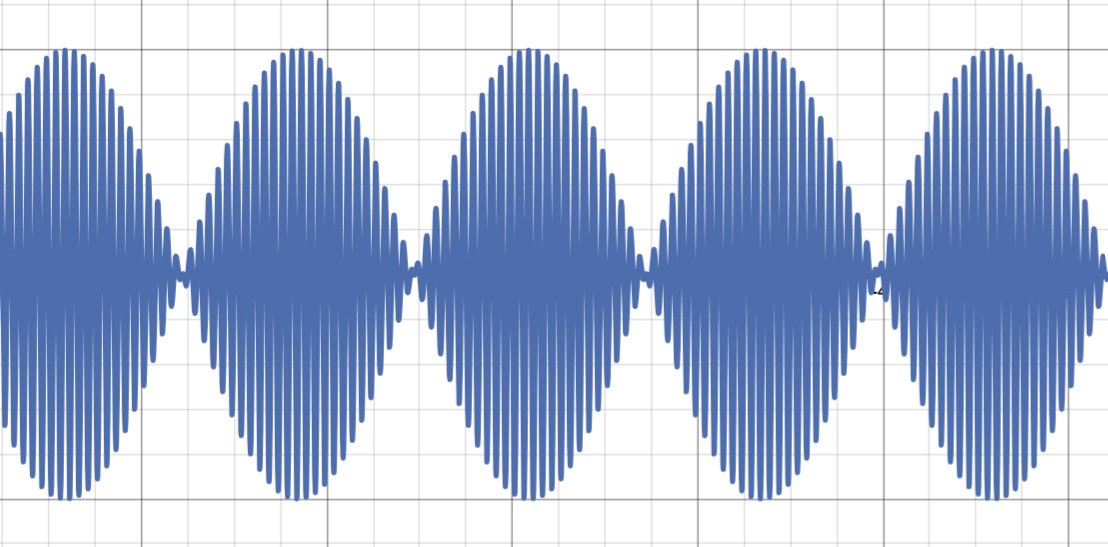}
    \label{cfo-a02}}
     \hfill
\subfloat[CFO = 0.3Hz]{
    \includegraphics[width=\linewidth,height=0.22\linewidth]{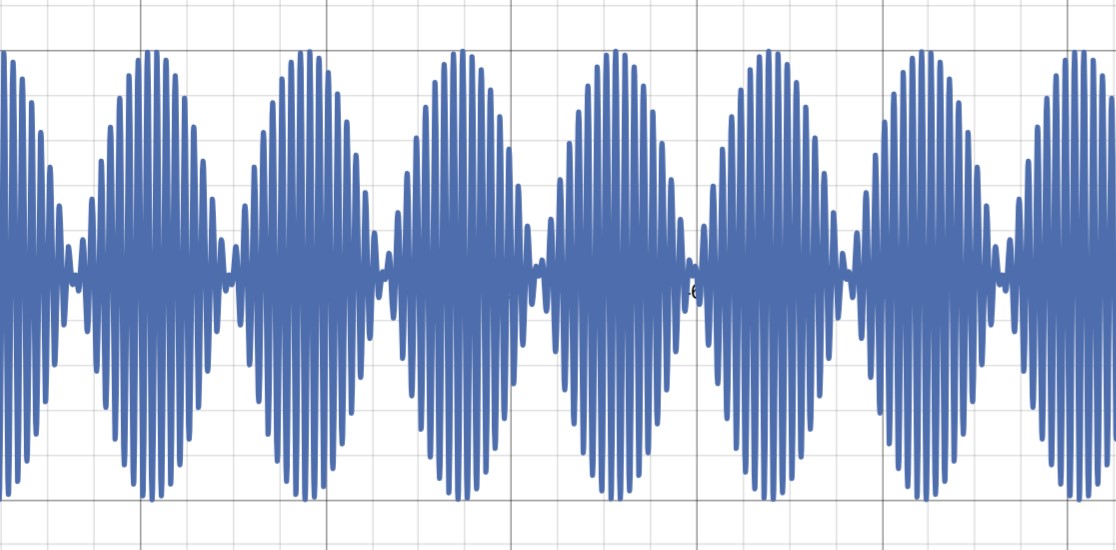}
    \label{cfo-a05}}
    \caption{CFO impact illustrations when $A(t)=\sin(20\pi t)$ and $\phi(t)=\pi/8$ under different values of CFO.} 
    \label{cfo-analytic} 
    \end{minipage}
\end{figure}

Recall that a bandpass signal $s_b(t)=A(t)e^{j\phi(t)}$ modulated with a carrier frequency $f_c$ having a CFO $\Delta f$ yields a passband signal $x(t) = A(t) \cos(2\pi (f_c + \Delta f) t + \phi(t))$.
%
At the receiver side, the demodulated In-phase (I) signal, $y_I(t)$, can be written as $x(t) \cos(2\pi f_ct)$ or (when replacing $x(t)$ with its expression)
\begin{equation}\nonumber
   \frac{A(t)}{2}[\cos(2\pi (2f_c  + \Delta f) t + \phi(t) ) + \cos(2\pi \Delta ft + \phi(t))].
    \label{env}
\end{equation}

Note that the term $\cos(2\pi (2f_c  + \Delta f) t + \phi(t) )$ will be eliminated through bandpass filtering but the term $\cos(2\pi \Delta ft + \phi(t))$ will remain, which captures the carrier frequency offset (CFO), $\Delta f$, resulting from the mismatch in the transmitter's and receiver's oscillating frequencies.
When CFO is zero, the bandpass filtered signal becomes $y_I(t)=\frac{A(t)}{2} \cos(\phi(t))$. But when CFO is non-zero, the filtered signal $\frac{A(t)}{2}\cos(2\pi \Delta f t + \phi(t))$ is impacted by the value of $\Delta f$.
Fig.~\ref{cfo-analytic} depicts the CFO impact on the signal behavior using an (toy) example with $A(t)=\sin(20\pi t)$ and $\phi(t)=\pi/8$ under different CFOs.
%

%

\subsection{CFO Impact Analysis Through Hardware Testing}

\begin{figure}
     \subfloat[Modified FiPy device.\label{modified}]{%
       \includegraphics[width=0.233\textwidth, height = 0.23\textwidth]{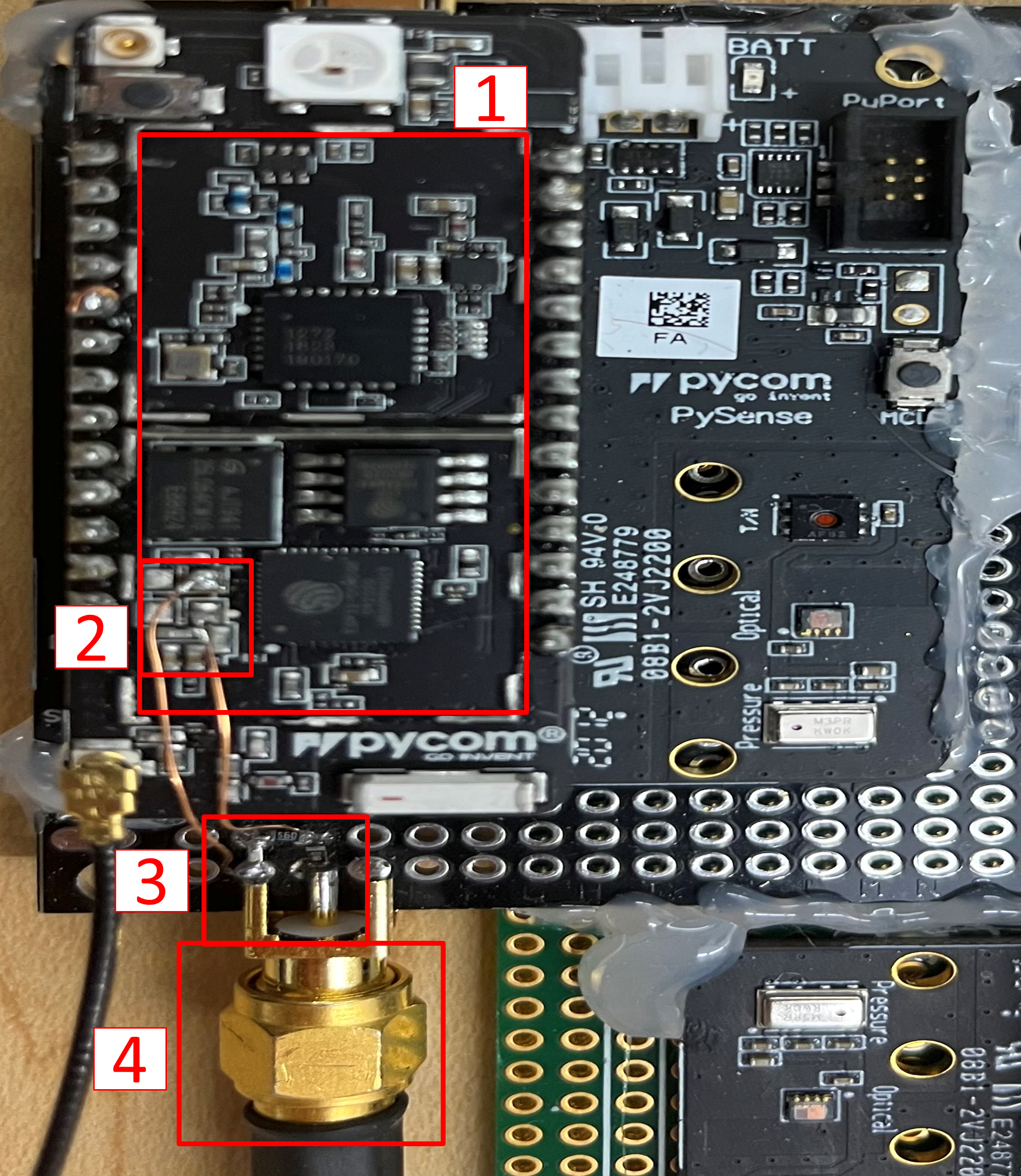}
     }
    \hspace{0.00001cm}
     \subfloat[Modified device setup.\label{mod_setup}]{%
       \includegraphics[width=0.233\textwidth, height = 0.23\textwidth]{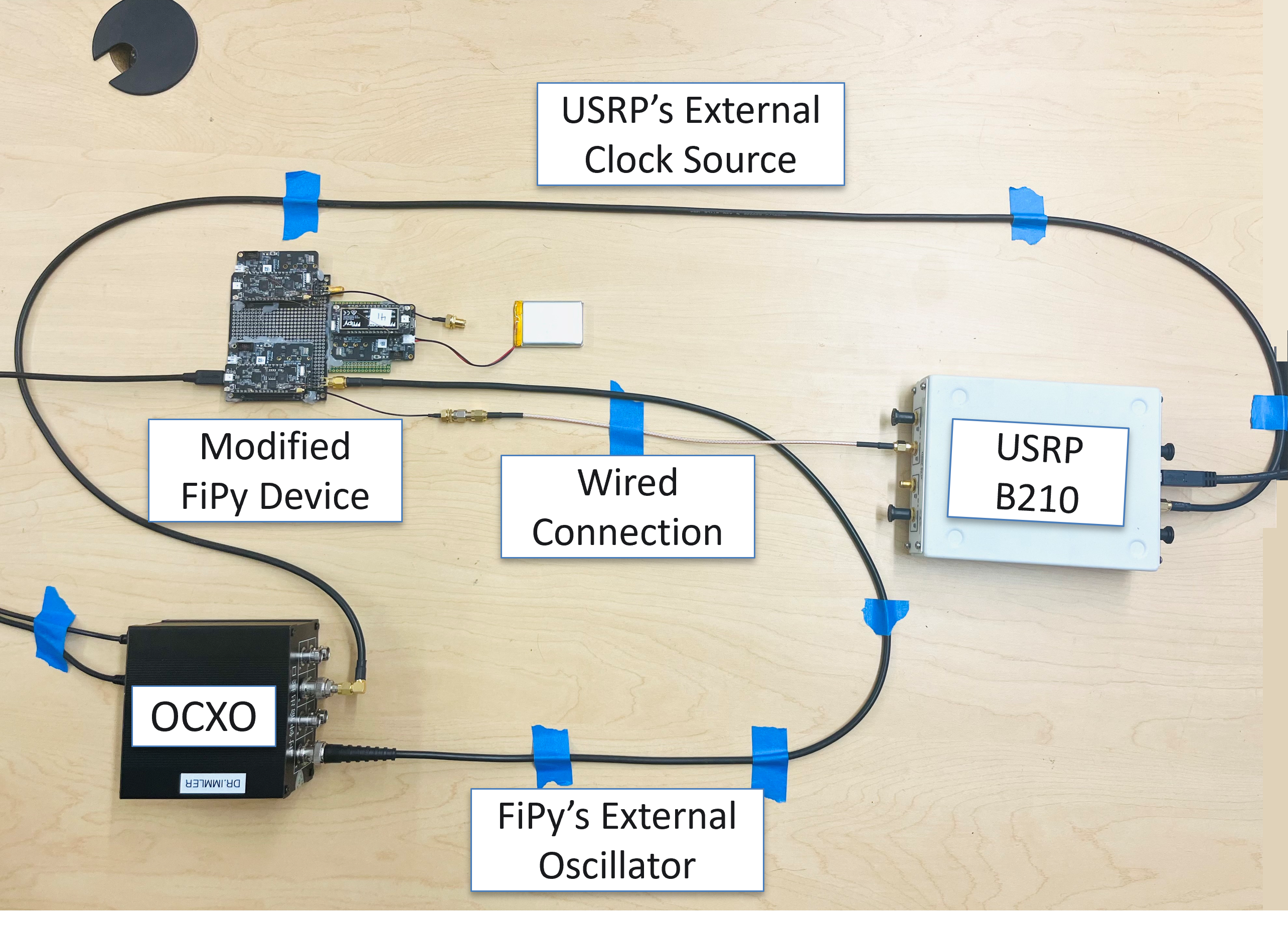}
     }
     \caption{The hardware setup: the modified FiPy device connected through an SMA cable to the USRP receiver.}
     \label{modified_setup}
\end{figure}

Now that we've analyzed and confirmed the relationship between the CFO impairments and the I/Q envelope behavior using Matlab simulations, we also confirm it here through hardware experimentation.
Our approach for this consisted of replacing the local oscillator of an off-the-shelf FiPy device with an external, high-end, oven-controlled crystal oscillator (OCXO) whose oscillating frequency is programmable and provides an order of magnitude of higher frequency accuracy and stability than the crystal oscillator hardware that came with the off-the-shelf FiPy devices. 
Specifically, referring to the picture of the modified board shown in Fig.~\ref{modified}, we first removed the EMI shield (red box $1$ in the figure) and the WiFi local crystal oscillator (red box $2$) that was found to be TST zTX crystal with a frequency of $40$MHz and frequency stability of $10$PPM. Next, we attached an SMA connector (red box $4$) to the side of the prototype board, which was connected to the output port of the programmable OCXO board through a BNC-SMA cable. A voltage divider (red box $3$) was created using surface-mounted resistors to adjust the voltage level of the external OCXO's signal, which was then connected to the pads of the SMA connector. Lastly, we connected the ground and signal outputs of the voltage divider to the corresponding pads of the oscillator in the FiPy board (red box $2$) using insulated copper wires. 
We then integrated the modified FiPy board into the wired connection setup, as depicted in Fig. \ref{mod_setup}. The OCXO, set to output an oscillating signal with a tunable frequency, is connected to the SMA connector in the prototype board to act as an external oscillator of the FiPy device. OCXO also outputs a $10$MHz-signal that is connected to the REF IN port of the USRP to act as an external clock source. These provided high-stability and better synchronization between the FiPy device and the USRP B$210$ receiver. We then set the modified FiPy board to transmit WiFi packets via the SMA wired connection, which were then sampled at $45$MSps.

\begin{figure}
\centering
\subfloat[CFO = 0]{%
   \includegraphics[width=.5\columnwidth,height=0.25\linewidth]{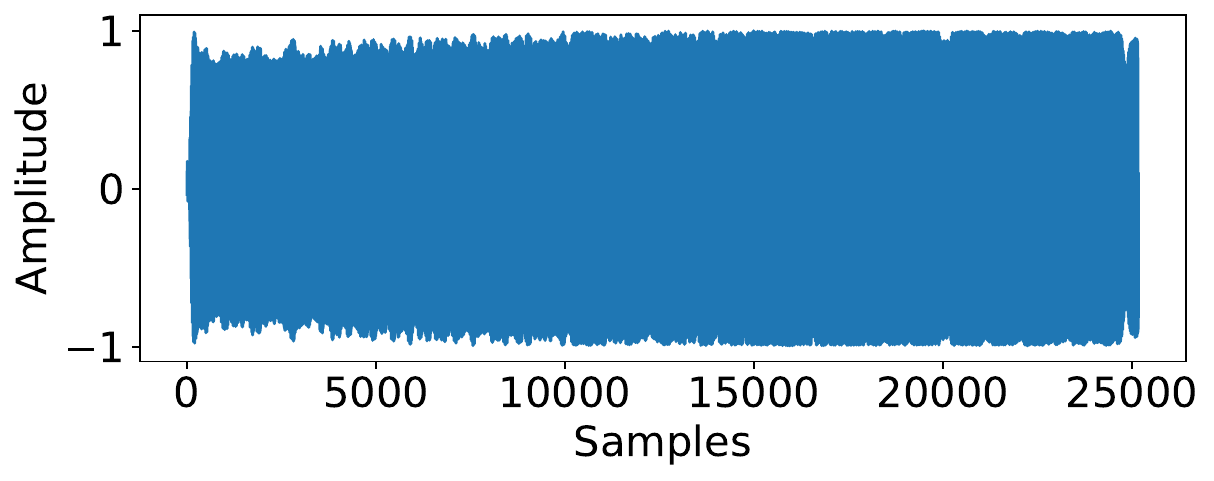}
   \label{fig:ocxo1}} 
\subfloat[CFO = 0.1 kHz]{%
   \includegraphics[width=.5\columnwidth,height=0.25\linewidth]{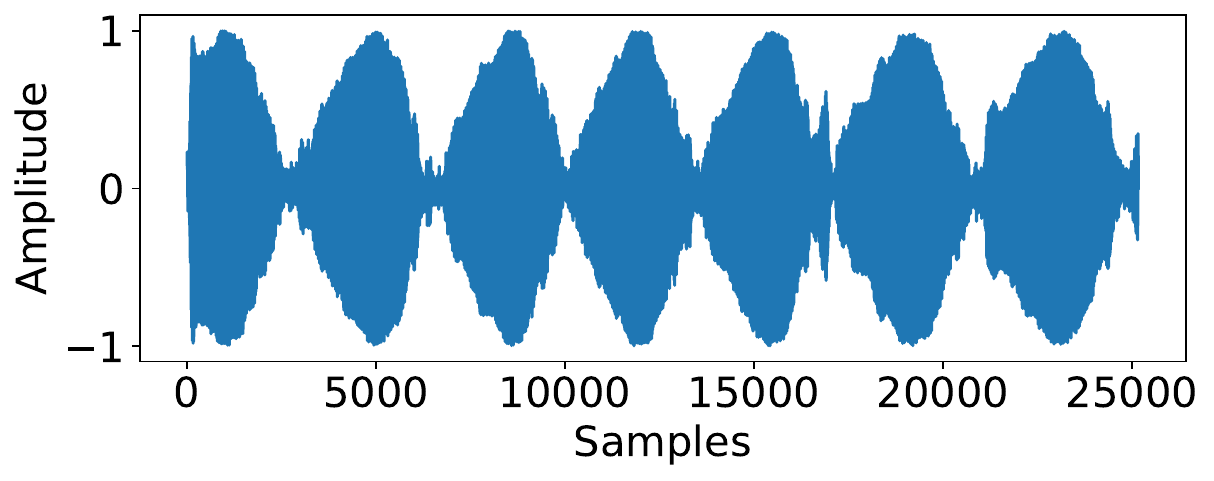}
   \label{fig:ocxo2}}\\
\subfloat[CFO = 0.5 kHz]{%
   \includegraphics[width=.5\columnwidth,height=0.25\linewidth]{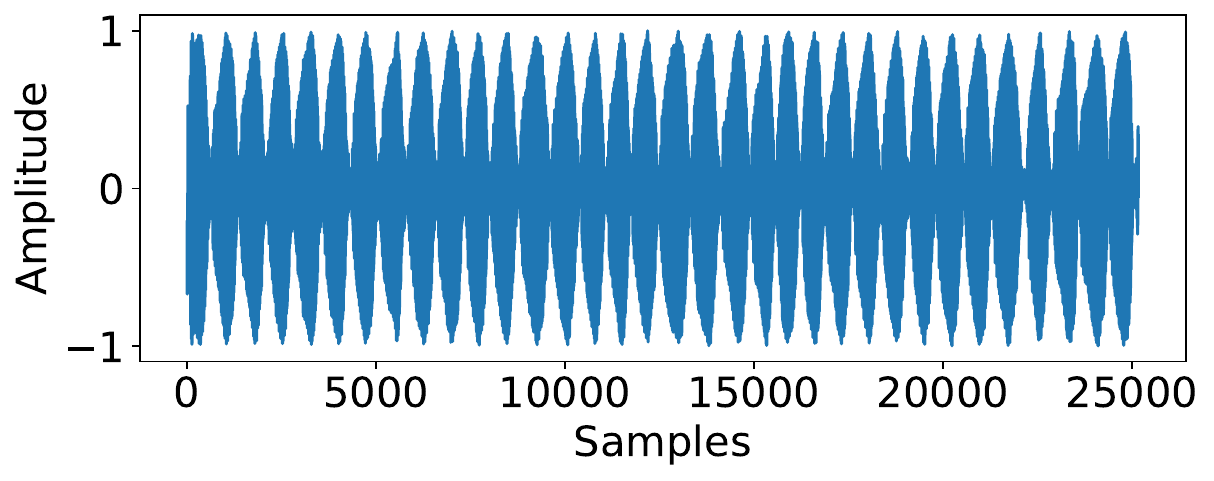}\label{subfig:ocxo3}
   \label{fig:ocxo3}} 
   \subfloat[CFO = 1 kHz]{%
   \includegraphics[width=.5\columnwidth,height=0.25\linewidth]{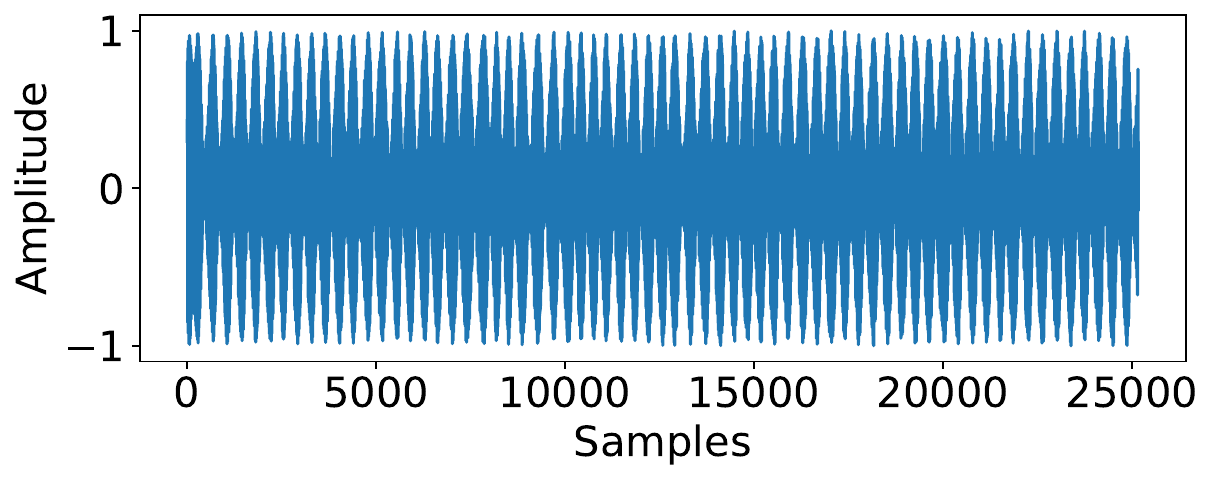}\label{subfig:ocxo4}
   \label{fig:ocxo4}}
\caption{The I signal component of a WiFi packet collected from the OCXO-enabled (modified) FiPy device.}
\label{fig:ocxo_all}
\end{figure}

Using this setup, we analyzed WiFi packets collected from the OCXO-enabled (modified) device while considering different oscillating frequency values, $f_{LO}=40$MHz (i.e., CFO = 0), $f_{LO}=39.9999$MHz (i.e., CFO = 0.1 kHz), $f_{LO}=39.9995$MHz (i.e., CFO = 0.5 kHz), and $f_{LO}=39.9990$MHz (i.e., CFO = 1 kHz), in order to evaluate the impact of CFO. Here, the USRP receiver's frequency is kept the same throughout all experiments.
We plotted in Fig.~\ref{fig:ocxo_all} the I component of the I/Q signals received from the modified FiPy device for the four  frequencies.  
Note that the time-domain I signal does exhibit a sinusoidal envelope behavior due to the presence of a CFO, and the number of humps in the envelope increases with the CFO value. Note also that when CFO = 0, the I signal yields a constant envelope.

In conclusion, our investigations confirm that this envelope behavior observed when using the original, low-performing crystal oscillators that came with the off-the-shelf Pycom devices is attributed to the CFO impairments. 
These investigations also confirm that devices with (even slightly) different oscillating frequencies yield different envelopes (i.e., different numbers of humps) of the received signals of the same transmitted packets, and such a difference across different devices could be leveraged to create and extract device-specific fingerprints.

\section{Conclusion}
\label{sec:conc}
This work studied the effect of hardware warm-up time on the behavior of the time-domain I/Q signals and the RF fingerprinting performance thereof. We revealed and explained the consequences of neglecting this hardware warm-up aspect when developing RF fingerprinting methods. Additionally, we demonstrated how the instability of the local oscillator during hardware warm-up impacts the behavior of I/Q signals and thus provided valuable insights into the observed decline in RF fingerprinting accuracy when the hardware stabilization aspects are not taken into account.
Finally, we released large WiFi 802.11b datasets containing captures collected before and after the stabilization of hardware components to allow others to perform further research investigations.

\section{Acknowledgment}
Many thanks to Dr. Richard Dorrance for  his constructive discussions and feedback and Ms. Nora Basha for her help with Matlab coding.


\bibliographystyle{IEEEtran}
\bibliography{IEEEexample}

\end{document}